\documentclass{article}

\PassOptionsToPackage{numbers, compress}{natbib}
\usepackage[eandd, preprint]{neurips_2026}

\usepackage[utf8]{inputenc} 
\usepackage[T1]{fontenc}    
\usepackage{hyperref}       
\usepackage{url}            
\usepackage{booktabs}       
\usepackage{amsfonts}       
\usepackage{nicefrac}       
\usepackage{microtype}      
\usepackage{xcolor}         
\usepackage{enumitem}
\usepackage{graphicx}
\usepackage{amsmath}
\usepackage{multirow}
\usepackage[table]{xcolor}

\usepackage{float}
\usepackage{threeparttable}
\usepackage{titlesec}  
\usepackage{appendix}  
\usepackage{tocloft}   
\usepackage{titletoc}
\usepackage{multicol}
\usepackage{multirow}
\usepackage{colortbl}
\usepackage{pifont}
\usepackage{graphicx}
\usepackage{subcaption}
\usepackage{etoolbox}
\usepackage{hyperref}
\usepackage{amsmath}
\usepackage{geometry}
\usepackage{booktabs}   
\usepackage{makecell}
\usepackage{enumitem}
\usepackage{amssymb}
\usepackage{textcomp}
\usepackage{array}
\usepackage{pifont}
\usepackage{float}      
\usepackage{placeins} 
\usepackage{capt-of} 

\usepackage{placeins} 

\title{CodegenBench: Can LLMs Write Efficient Code Across Architectures?}

\begin{document}

\author{
\textbf{Jie Li}$^{1,*}$,
\textbf{Wenzhao Wu}$^{2,*}$,
\textbf{Junqi Hu}$^{1,*}$,
\textbf{Qinrui Zheng}$^{3,*}$,
\textbf{Bowen Wu}$^{4}$, \\
\textbf{Juepeng Zheng}$^{1,3,\dagger}$ 
\textbf{Yutong Lu}$^{1,3}$,
\textbf{Haohuan Fu}$^{3,5}$, \\
$^{1}$Sun Yat-sen University, 
$^{2}$National Supercomputing Center in Wuxi, \\
$^{3}$National Supercomputing Center in Shenzhen \\
$^{4}$University of the Chinese Academy of Sciences,
$^{5}$Tsinghua Shenzhen International Graduate School \\
$^{*}$Equal contribution, $^{\dagger}$Corresponding author. 
}
\maketitle

\begin{abstract}
While large language models (LLMs) have been extensively evaluated on code generation tasks for general-purpose programming and GPU-accelerated environments (e.g., PyTorch, CUDA), their capabilities in CPU-oriented high-performance computing (HPC) across diverse architectures remain underexplored. To bridge this gap, we introduce \textbf{CodegenBench}, a comprehensive benchmark suite designed to evaluate the generation of efficient parallel code across three distinct hardware platforms: x86\_64, Sunway, and Kunpeng. Our benchmark comprises 106 standard Basic Linear Algebra Subprograms (BLAS) routines establishing a fundamental baseline, alongside 20 specialized computational kernels adapted for each of the unique supercomputing architectures (LeetSunway and LeetKunpeng). Our extensive evaluation reveals that while state-of-the-art LLMs can generate optimized code for ubiquitous architectures like x86\_64, they exhibit significant performance degradation on domain-specific architectures with limited public documentation and training data, highlighting critical limitations in cross-platform generalization. Furthermore, our analysis of factors influencing code quality such as implementation length and task complexity indicates that current LLMs are most effective for moderately difficult problems requiring concise code snippets. We open-source our dataset and automated evaluation infrastructure to facilitate future research in LLM-driven high-performance code generation. The resources are available at \url{https://anonymous.4open.science/r/CodegenBench-EDE1/} and \url{https://anonymous.4open.science/r/CodegenBenchDataset-2551/}.
\end{abstract}

\section{Introduction}

The rapid advancement of Large Language Models (LLMs) has catalyzed a paradigm shift in software engineering, enabling highly efficient generation of standard code components, ranging from web applications to interactive scripts~\cite{guo2024deepseek, zhu2024deepseek, allal2023santacoder, li2023starcoder,bai2025qwen3,sobo2025evaluating,zeng2026glm}. General-purpose models like GPT~\cite{brown2020language}, Gemini~\cite{team2024gemini}, and Claude~\citep{priyanshu2024ai}, alongside domain-specific tools such as GitHub Copilot~\cite{chen2021evaluating}, have fundamentally transformed developer productivity. Today, AI-assisted programming has become a staple, consistently delivering fast and functionally correct code.

Despite these remarkable strides, leveraging LLMs to produce highly optimized, performance-critical code, particularly in HPC contexts, remains a formidable challenge. While recent initiatives have begun to explore this frontier, they predominantly focus on specific, widely-used domains. For instance, KernelBench~\cite{ouyang2025kernelbench} evaluates whether LLMs can generate efficient CUDA kernels compared to standard PyTorch libraries, and Effibench~\cite{huang2024effibench} assesses the execution efficiency of LLM-generated code on LeetCode problems. \citet{mukunoki2025performance} also provided preliminary insights into LLMs' capacity for generating high-performance BLAS routines.

However, existing evaluations are largely constrained by their focus on mainstream hardware platforms and AI-centric software stacks. Prior frameworks predominantly target GPU environments~\cite{ouyang2025kernelbench, li2025tritonbench, wen2025multikernelbench} and Python-based tasks~\cite{austin2021program, chen2021evaluating, huang2024effibench, NEURIPS2023_43e9d647, qiu2024efficient}, often prioritizing low-precision, high-throughput computation tailored for AI models and NVIDIA CUDA ecosystems. This narrow focus leaves a critical gap in assessing LLMs' capabilities for general-purpose scientific computing on non-CUDA, heterogeneous CPU architectures. In demanding HPC scenarios, the core challenge stems less from the inherent algorithmic complexity of the tasks and significantly more from the intricate hardware-aware adaptation required. Specifically, generating code in such environments entails three primary challenges: \textit{(i) Portability}: porting existing baseline code (e.g., x86) while maintaining or improving performance; \textit{(ii) Architecture-awareness}: leveraging diverse and unique architectural features requiring specialized hardware knowledge; \textit{(iii) Data Scarcity}: overcoming the scarcity of high-quality, open-source training data for proprietary or domain-specific platforms.

\begin{figure}[t]
  \centering
    \includegraphics[width=\linewidth]{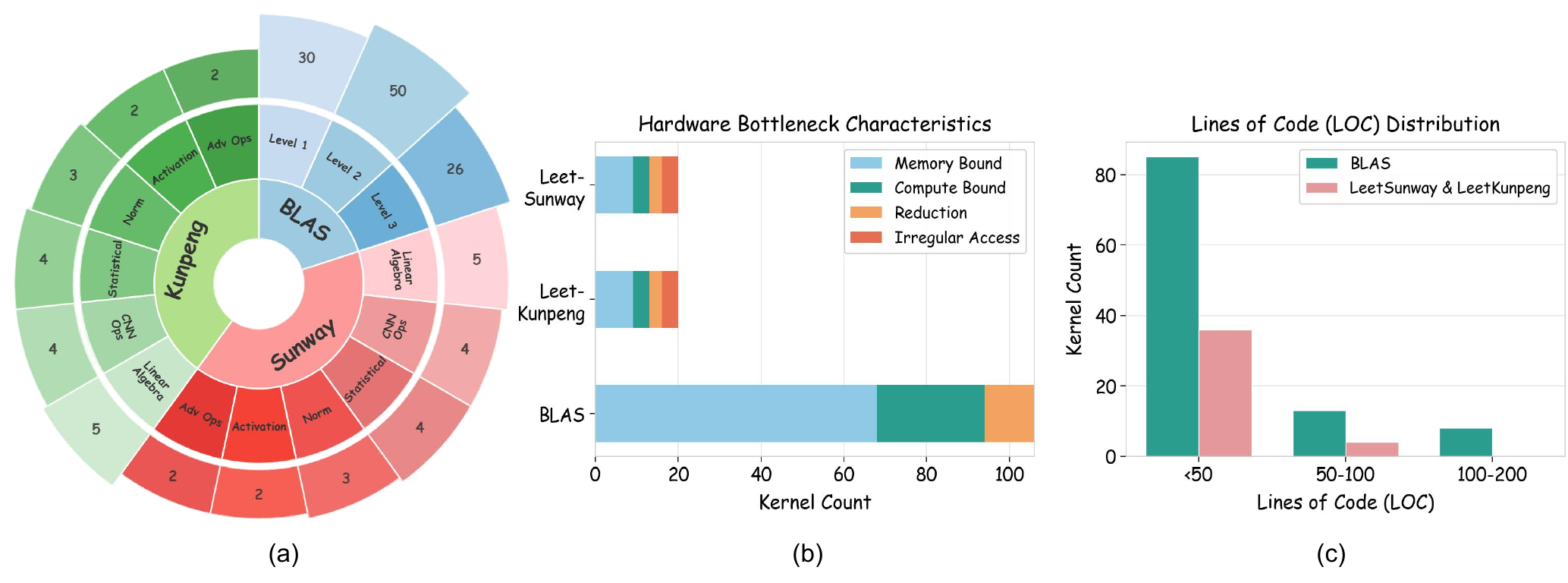}
    \vspace{-2em}
    \caption{Overall statistic of CodegenBench.}
    \label{fig:stat}
    \vspace{-1.5em}
\end{figure}

To address these challenges, we introduce \textbf{CodegenBench}, a comprehensive evaluation framework equipped with reference implementations and platform-specific tasks from LLMs to rigorously assess capacity for generating scalable, hardware-aware code. The benchmark explicitly targets performance-critical tasks across three domains. As presented in Fig.~\ref{fig:stat}, the BLAS suite covers standard numerical routines, establishing a fundamental baseline. In contrast, the LeetKunpeng and LeetSunway subsets strictly demand tailored code generation for the Kunpeng~\cite{xia2021kunpeng, afanasyev2021evaluating} and Sunway~\cite{fu2016sunway} architectures which are non-x86, non-CUDA, CPU-focused supercomputing platforms. Our key contributions are:

\begin{figure}
  \centering
    \includegraphics[width=\linewidth]{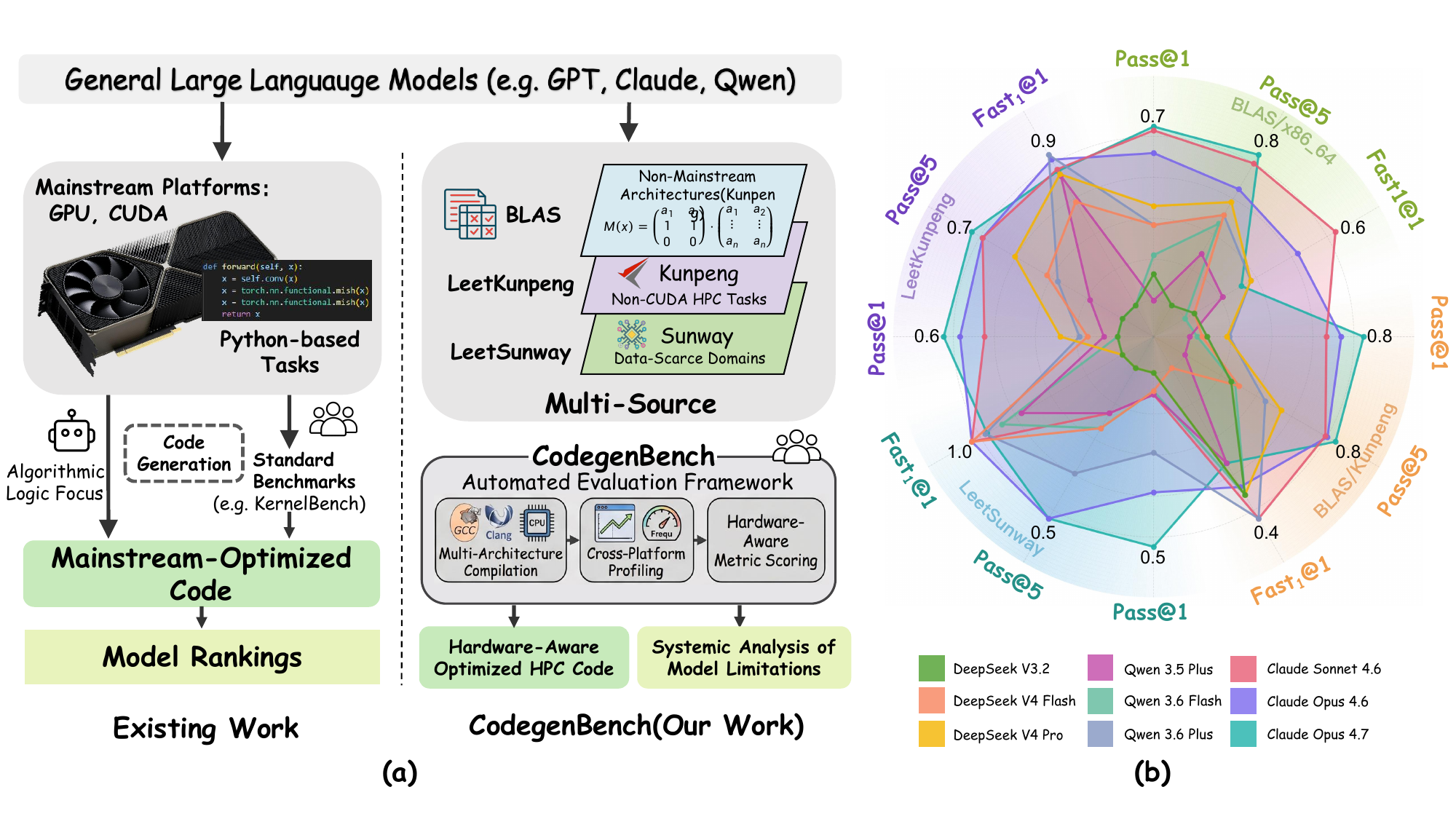}
    \vspace{-2em}
    \caption{Comparison of our framework with existing paradigms, and its comprehensive performance radar on the CodegenBench benchmark. Existing benchmarks mainly target mainstream platforms like GPU, CUDA. CodegenBench aims to target x86\_64, Sunway and Kunpeng architecture with hand-written test case for these platforms.}
    \label{fig:CodegenBenchComparison}
    \vspace{-1.5em}
\end{figure}

\begin{itemize}[leftmargin=*, itemsep=1em, parsep=0pt, topsep=0pt]
\item We develop CodegenBench, a highly extensible, automated evaluation framework that streamlines the end-to-end life-cycle of LLM-driven code generation, cross-platform compilation, and performance profiling.
\item We introduce multi-architecture benchmark data comprising benchmark tasks that diverge from standard AI-centric workloads. This includes a comprehensive set of BLAS routines and frequently used computational operators (LeetKunpeng and LeetSunway), complete with baseline CPU reference implementations.
\item We conduct an extensive empirical evaluation of top-tier open-source and closed-source LLMs across three architecturally distinct hardware platforms. Our findings systematically expose current model limitations in hardware-specific optimization and identify key factors contributing to performance degradation in cross-architecture code generation.
\end{itemize}

\newcommand{\cmark}{$\checkmark$}
\newcommand{\xmark}{$\times$}

\begin{table}[t]
\centering
\caption{Comparison of benchmark datasets across scalability, portability and multi-architecture. Pass@k denotes using iteration to evaluate correctness; Speedup denotes evaluating the improvement of parallel code compared to sequential code; HPC denotes containing HPC benchmark; Sunway denotes containing Sunway architecture; Kunpeng denotes containing Kunpeng architecture; \cmark: fully supported; \xmark: not supported.}
\label{tab:benchmark-comparison}
\scriptsize
\begin{tabular}{lcccccccc}
\toprule
Dataset & HPC & Scalable & Portable & Multi-architecture & Pass@k & Speedup & Sunway & Kunpeng \\
\midrule
KernelBench~\citep{ouyang2025kernelbench}      & \xmark & \xmark & \xmark & \xmark & \xmark & \cmark & \xmark & \xmark \\
MultiKernelBench~\citep{wen2025multikernelbench} & \xmark & \cmark & \cmark & \cmark & \xmark & \xmark & \xmark & \xmark \\
HosNa~\citep{bavarsad2021hosna}            & \xmark & \xmark & \cmark & \cmark & \xmark & \xmark & \xmark & \xmark \\
\midrule
BLAS~\citep{duff2002overview}             & \cmark & \cmark & \cmark & \cmark & \cmark & \cmark & \xmark & \xmark \\
PCEBench~\citep{chen2025pcebench}         & \cmark & \cmark & \cmark & \xmark & \cmark & \cmark & \xmark & \xmark \\
CUDABench~\citep{zhu2026cudabench}        & \cmark & \xmark & \xmark & \xmark & \xmark & \xmark & \xmark & \xmark \\
HeteroBench~\citep{tian2025heterobench}      & \cmark & \cmark & \cmark & \cmark & \xmark & \xmark & \xmark & \xmark \\
DRB-ML~\citep{chen2023data}           & \cmark & \xmark & \xmark & \xmark & \xmark & \xmark & \xmark & \xmark \\
ParEval~\citep{nichols2024can}          & \cmark & \cmark & \xmark & \xmark & \cmark & \xmark & \xmark & \xmark \\
LLM4VV~\citep{munley2024llm4vv}           & \cmark & \xmark & \cmark & \cmark & \xmark & \xmark & \xmark & \xmark \\
\midrule
\rowcolor{gray!15}
\textbf{CodegenBench} (Ours)     & \cmark & \cmark & \cmark & \cmark & \cmark & \cmark & \cmark & \cmark \\
\bottomrule
\end{tabular}
\vspace{-2em}
\end{table}

\section{Related Work}
\subsection{Large Language Models for Programming}
With the evolution of LLMs, programming, being a linguistic and logistic work, is covered within the capabilities of modern LLMs. Codex~\citep{chen2021evaluating} is one of the very first LLMs specifically fine-tuned for code generation, while later models like Code Llama~\citep{roziere2023code}, StarCoder~\citep{li2023starcoder}, SantaCoder~\citep{allal2023santacoder} have further advanced in the field. With more attention paid to the performance of programming, more and more models push the code generation capabilities further forward, such as Gemini~\citep{team2024gemini} and Claude~\citep{priyanshu2024ai}. To keep pace with the development of these leading models, Chinese companies have studied and released their own models, including DeepSeek~\citep{guo2024deepseek}, Qwen~\citep{yang2025qwen2}, MiniMax~\citep{chen2025minimax} and GLM~\citep{zeng2025glm} showing promising competitiveness. Currently, these models are continually updated, showing incredible ability to solve programming problems~\citep{bai2025qwen3,sobo2025evaluating,zeng2026glm}.

However, although LLMs show remarkable capabilities in sequential code generation, they are not good at generating parallel code. Chen et al.~\citep{chen2025pcebench} show that when dealing with OpenMP case, the passed samples are less than 70\% in the targeted sources even though iterating several times. The MPI case is even worse. And the research demonstrates that generating code for HPC benchmarks is more difficult than sequential problems. So it is vital to come up with a new method to help LLMs optimize the capability to generate parallel code especially in HPC benchmarks.

\subsection{LLM for Parallel Coding}

When dealing with code generation tasks, the precision and fulfillment of user intentions are of importance due to the prominent character of human preferences different from open text generation~\citep{ouyang2022training}. Although traditional evaluation metrics such as BLEU~\citep{papineni2002bleu} and ROUGE~\citep{lin2004rouge} are not able to cope with this situation, several recently proposed benchmarks including MBXP~\citep{athiwaratkun2022multi}, CodeContests~\citep{li2022competition}, and DS-1000~\citep{lai2023ds} exhibit good performance. But they are designed for sequential code not applicable to parallel code. Recent years have seen growing effort in evaluating the performance of parallel and High Performance Computing (HPC) codes, especially with AI's rapid development. Using various prompt engineering techniques~\citep{munley2024llm4vv} evaluate the capabilities of LLMs by generating tests and using these tests to verify compiler implementations for parallel OpenACC code means starting the construction of parallel code generation benchmarks. DRB-ML dataset~\citep{chen2023data} is then proposed by exploring a novel LLM-based data race detection approach. As high-performance computing kernels become important in application, the capabilities of LLMs to generate HPC kernels are evaluated~\citep{valero2023comparing} afterwards, especially CUDA case~\citep{zhu2026cudabench}. To make a comprehensive evaluation and comparison of the ability of LLMs to generate parallel code, ParEval benchmark~\citep{nichols2024can} consisting of prompts representing 420 different coding tasks related to scientific and parallel computing comes into sight. KernelBench is introduced later~\citep{ouyang2025kernelbench} to evaluate LMs' ability to write fast and correct kernels as efficient GPU kernels are crucial for building performant machine learning architectures. 

However, these benchmarks focus only on a single architecture~\citep{chen2023data,chen2025pcebench,nichols2024can,ouyang2025kernelbench,zhu2026cudabench} that is not compatible when dealing with other architectures. Even though some recently proposed benchmarks such as HeteroBench~\citep{tian2025heterobench}, HosNa~\citep{bavarsad2021hosna} and MultiKernelBench~\citep{wen2025multikernelbench} concerned about multi-architecture including CPU, GPU, NPU, TPU and FPGA, the specific architectures used by supercomputers such as Sunway and Kunpeng that are of importance to have criteria in are not included. In this paper, we introduce a new benchmark called CodegenBench to test the performance of generated parallel codes on different architectures and platforms to fill the gap. Table 1 shows the strength of our benchmark compared with others.

\vspace{-0.5em}
\section{Benchmark Composition}
\vspace{-0.5em}
In this section, we present a comprehensive overview of the benchmark composition utilized in our evaluation framework. The bench is meticulously constructed and divided into three primary components: one extensive BLAS part targeting general-purpose computing platforms, and two specialized supercomputer parts, LeetSunway and LeetKunpeng, targeting the Sunway and Kunpeng architectures, respectively. This diverse composition allows for a robust assessment across both standard mathematical libraries and heterogeneous computing environments.

\subsection{BLAS}
The BLAS portion of the bench acts as a fundamental baseline, comprising 106 distinct BLAS subroutines spanning its standard three levels. The inherent hierarchical structure of BLAS naturally presents varying degrees of optimization difficulty. Because the operational complexity and memory access patterns differ significantly from Level 1 to Level 3, the requirements for generating optimized code scale accordingly. This built-in gradient of difficulty makes the BLAS an excellent baseline for evaluating a model’s capabilities in both basic code generation and advanced algorithmic optimization. Furthermore, each individual subroutine is evaluated under various parameter combinations such as varying matrix dimensions, strides, and data types to ensure that the performance, accuracy, and edge-case handling of the LLM-generated code are comprehensively assessed.

\subsection{Sunway}

\textbf{Sunway Supercomputer} operates on SW26010 many-core processors. This advanced hardware features a highly specific and complex topology, consisting of four distinct Core Groups (CGs) per processor chip. Within each Core Group, there is one Management Processing Element (MPE, or master core) designed for control and task scheduling, alongside a cluster of 64 Computing Processing Elements (CPEs, or slave cores) dedicated to intensive parallel data processing~\citep{fu2016sunway}. Unlike standard x86 architectures, the unique distributed memory and master-slave core paradigm of the SW26010 processor demands in-depth architectural research, explicit data routing mechanisms, and extensive low-level tuning knowledge to manually write and optimize computing kernels. Consequently, generating both functionally correct and high-performance codes in this highly specialized, non-standard scenario presents a formidable challenge for contemporary LLMs, severely testing capacity to comprehend and output code for constrained processing environments.

\textbf{LeetSunway} is a meticulously curated problem set consisting of exactly 20 typical and frequently utilized computation kernels tailored specifically for the Sunway architecture. The selected kernels encompass essential operations across various computational domains. For each specific test case within this suite, the generated code must fully leverage the aforementioned master-slave core paradigm, applying appropriate thread synchronization and memory access strategies to maximize the overall hardware performance of the target platform. Given the inherent scarcity of publicly available training data and documentation regarding the Sunway architecture's proprietary instruction sets, these 20 diverse benchmark problems serve as an exceptional challenging test to evaluate capabilities in specialized code optimization, parallel computing logic, and novel architecture porting.

\subsection{Kunpeng}

\textbf{Kunpeng} is another facet of heterogeneous systems, which represents a distinctive alternative architectural paradigm compared to the highly proprietary Sunway architecture. The Kunpeng processor family utilizes an ARM-based underlying design~\cite{xia2021kunpeng}, enriched with advanced HPC features such as On-die High Bandwidth Memory (HBM) and powerful parallel processing capabilities like Scalable Vector Extension (SVE) and Scalable Matrix Extension (SME). Although Kunpeng and general ARM servers historically exhibit lower mainstream adoption in conventional HPC centers when directly compared to the ubiquitous x86 architecture, the standardized ARM-based instruction set means that there are significantly more relevant, high-quality code examples and optimization tutorials actively circulating in public code repositories compared to Sunway. Therefore, systematically evaluating LLMs on this architecture provides crucial comparative insights into their code generation capabilities when dealing with moderately represented but standard architectures, bridging the gap between hyper-specialized (Sunway) and ultra-common (x86) training distributions.

\textbf{LeetKunpeng} benchmark mirrors its Sunway counterpart exactly in terms of the algorithmic tasks required. Specifically, the 20 computing kernels in LeetKunpeng are identical to those formulated for LeetSunway. Despite having the same problem set, LeetKunpeng focuses on utilizing the unique hardware features of the Kunpeng chip. For each test case, the expectation is to evaluate the LLM's capacity to generate functional codes that properly target the ARM-based Kunpeng architecture.  

\vspace{-0.5em}
\section{CodegenBench Framework}
\begin{figure}[t]
\centering
\label{pipeline}
\includegraphics[width=\linewidth]{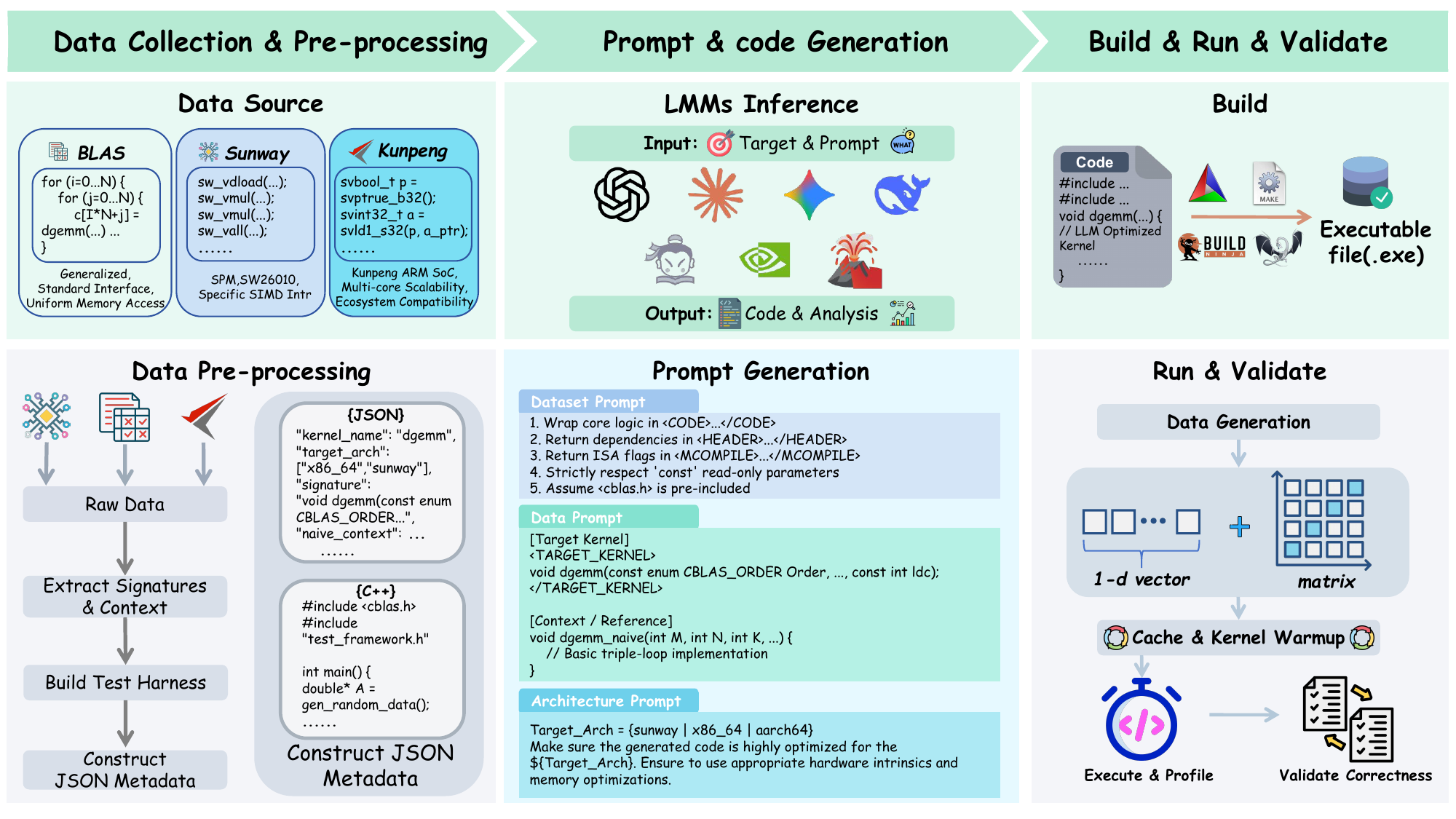}
\vspace{-2em}
\caption{Overall pipeline for CodegenBench. The pipeline automates evaluation.}
\vspace{-1.5em}
\end{figure}

\vspace{-0.5em}
In this section, we comprehensively introduce the automated framework which we developed and utilized for our evaluation methodology. As presented in Fig.~\ref{pipeline}, to ensure rigorous, reproducible, and scalable testing across various scenarios, the framework is methodically divided into four distinct stages. A key design principle of this pipeline is its high degree of adaptability; each stage can be extensively customized, either through configuration files or slight modifications to the underlying code, providing the necessary flexibility to seamlessly support varied and complex benchmark data structures without requiring a complete overhaul of the testing infrastructure.

\textbf{Pre-processing.}
In the initial pre-processing stage, the framework's primary objective is to establish a pristine and deterministic environment for each evaluation. It begins by parsing the provided metadata and executing a series of preliminary commands before any generation or build processes begin. These preliminary tasks typically include cleaning up temporary files and the legacy code base to initialize working directories, and setting up necessary environment variables. By resetting the workspace, this step helps with consistency and fairness across parallel or sequential evaluations, ensuring that the initial environments remain practically identical and free from cross-contamination.

\textbf{Prompt Generation and Code Generation.}
Following environmental setup, the framework transitions into the prompt generation and code generation stage. Here, the system aggregates context by extracting specific instructions from the metadata, alongside combining source code snippets, documentation, or problem descriptions from related files. These scattered pieces of information are seamlessly woven together to construct a comprehensive and well-formatted prompt. Subsequently, the framework issues an API call to the target LLM to elicit the requested structural output, such as specific code blocks. Furthermore, both the prompt templates and the metadata schemas can be dynamically adjusted or enhanced, empowering the framework to handle increasingly complex benchmark data or specialized domain-specific requirements with minimal friction.

\textbf{Build.}
Once the preliminary setup and code generation have been completed, the pipeline moves on to the build stage. The framework automatically parses and executes the precise compilation directives specified within the setup metadata. Generally, this phase involves invoking standard industry compilers or automatic build tools such as CMake and Makefile to compile the LLM-generated code alongside the benchmark harness. The reliability of this stage serves as an initial litmus test for the model's output. If the build process fails, whether due to syntax errors, hallucinated function calls, or missing dependencies, it essentially indicates that the LLM-generated code falls short of forming a fundamentally executable program. Such failures are meticulously logged as they reflect a tangible limitation in the model's coding capabilities and logical coherence.

\textbf{Run and Validation.}
The final stage encompasses the actual execution and empirical validation of the compiled programs. The framework launches the executables, injecting the runtime parameters defined in the metadata. Simultaneously, execution logs, standard outputs, and error streams are systematically collected. The results are then compared against expected outcomes using tolerance-based numerical comparisons. This thorough data collection is instrumental for conducting granular post-hoc analysis, evaluating algorithmic correctness, and measuring overall performance metrics.

\vspace{-1em}
\section{Evaluation}
\vspace{-1em}

\subsection{Experiment Setup}
\label{sec:exp_setup}
\vspace{-0.5em}

\textbf{Metrics.} We chose three metrics in our evaluation: $Pass@1$, $Pass@5$ and $Fast_1@1$. These metrics will reflect the ability to generate correct and efficient code for LLMs (details in Appendix \ref{app:metric}).

\textbf{Evaluation Platform.} We conducted the experiments on three platforms. The x86\_64 platform is equipped with Xeon Platinum 8488C Processor. The Kunpeng platform is equipped with latest Kunpeng processor, with SVE/SME and On-die HBM. The Sunway platform is the Sunway Taihulight supercomputer, with SW26010 many-core processors. More details are provided in Appendix \ref{app:evaluation_environments}.

\textbf{Evaluation Models.} We evaluated a total of 9 LLMs using CodegenBench, including 5 open-sourced models and 4 close-sourced models. The open-sourced models include DeepSeek V3.2~\citep{liu2025deepseek}, DeepSeek V4 Flash~\citep{deepseek_v4_flash}, DeepSeek V4 Pro~\citep{deepseek_v4_pro}, Qwen3.5 Plus~\citep{qwen_35} and Qwen 3.6 Flash~\citep{qwen_36_35b}. The close-sourced models include Claude Sonnet 4.6~\citep{claude_sonnet_46}, Claude Opus 4.6~\citep{claude_opus_46}, Claude Opus 4.7~\citep{claude_opus_47} and Qwen 3.6 Plus~\citep{qwen_36}. More details are provided in Appendix \ref{app:evaluation_models}.

\begin{table}[t]
\centering
\caption{Evaluation results on BLAS-x86\_64 and LeetSunway. Best results across all models are shown in \textbf{bold}, and second-best results are \underline{underlined}.}
\label{tab:result_x86_sunway}
\small
\begin{tabular}{lcccccc}
\toprule
\multirow{2}{*}{Model}
& \multicolumn{2}{c}{$Pass@1$}
& \multicolumn{2}{c}{$Pass@5$}
& \multicolumn{2}{c}{$Fast_1@1$} \\
\cmidrule(lr){2-3} \cmidrule(lr){4-5} \cmidrule(lr){6-7}
& BLAS-x86 & LeetSW
& BLAS-x86 & LeetSW
& BLAS-x86 & LeetSW \\
\midrule
\rowcolor{gray!15}
\textbf{Closed-source} & & & & & & \\
Claude Sonnet 4.6~\citep{claude_sonnet_46} & \underline{0.70} & 0.06 & \underline{0.79} & 0.15 & \textbf{0.62} & \textbf{1.00} \\
Claude Opus 4.6~\citep{claude_opus_46}   & 0.67 & \underline{0.33} & 0.77 & \textbf{0.50} & \underline{0.58} & \textbf{1.00} \\
Claude Opus 4.7~\citep{claude_opus_47}   & \textbf{0.74} & \textbf{0.48} & \textbf{0.85} & \textbf{0.50} & 0.52 & 0.90 \\
Qwen 3.6 Plus~\citep{qwen_36}     & 0.48 & 0.22 & 0.71 & \underline{0.35} & 0.53 & \underline{0.91} \\
\rowcolor{gray!15}
\textbf{Open-source} & & & & & & \\
DeepSeek V3.2~\citep{liu2025deepseek}    & 0.35 & 0.00 & 0.50 & 0.00 & 0.47 & N/A  \\
DeepSeek V4 Flash~\citep{deepseek_v4_flash} & 0.48 & 0.05 & 0.71 & 0.20 & 0.47 & \textbf{1.00} \\
DeepSeek V4 Pro~\citep{deepseek_v4_pro}   & 0.53 & 0.00 & 0.74 & 0.00 & 0.53 & N/A  \\
Qwen 3.6 Flash~\citep{qwen_36_35b}    & 0.40 & 0.00 & 0.69 & 0.00 & 0.46 & N/A \\
Qwen 3.5 Plus~\citep{qwen_35}     & 0.28 & 0.00 & 0.62 & 0.00 & 0.50 & N/A \\
\bottomrule
\end{tabular}
\vspace{-0.5em}
\end{table}

\begin{table}[t]
\centering
\caption{Evaluation results on BLAS-Kunpeng and LeetKunpeng. Best results across all models are shown in \textbf{bold}, and second-best results are \underline{underlined}.}
\label{tab:result_kunpeng}
\small
\begin{tabular}{lcccccc}
\toprule
\multirow{2}{*}{Model}
& \multicolumn{2}{c}{$Pass@1$}
& \multicolumn{2}{c}{$Pass@5$}
& \multicolumn{2}{c}{$Fast_1@1$} \\
\cmidrule(lr){2-3} \cmidrule(lr){4-5} \cmidrule(lr){6-7}
& BLAS-KP & LeetKP
& BLAS-KP & LeetKP
& BLAS-KP & LeetKP \\
\midrule
\rowcolor{gray!15}
\textbf{Closed-source} & & & & & & \\
Claude Sonnet 4.6~\citep{claude_sonnet_46} & 0.61 & 0.49 & 0.80 & \underline{0.65} & \textbf{0.39} & 0.84 \\
Claude Opus 4.6~\citep{claude_opus_46}   & \underline{0.67} & \underline{0.58} & \underline{0.81} & \underline{0.65} & 0.35 & \underline{0.90} \\
Claude Opus 4.7~\citep{claude_opus_47}   & \textbf{0.76} & \textbf{0.64} & \textbf{0.85} & \textbf{0.70} & 0.32 & 0.83 \\
Qwen 3.6 Plus~\citep{qwen_36}     & 0.22 & 0.14 & 0.50 & 0.30 & \textbf{0.39} & \textbf{0.93} \\
\rowcolor{gray!15}
\textbf{Open-source} & & & & & & \\
DeepSeek V3.2~\citep{liu2025deepseek}     & 0.13 & 0.00 & 0.33 & 0.00 & \underline{0.36} & N/A  \\
DeepSeek V4 Flash~\citep{deepseek_v4_flash} & 0.11 & 0.11 & 0.37 & 0.35 & 0.20 & 0.64 \\
DeepSeek V4 Pro~\citep{deepseek_v4_pro}   & 0.21 & 0.21 & 0.58 & 0.50 & \underline{0.36} & 0.81 \\
Qwen 3.6 Flash~\citep{qwen_36_35b}    & 0.09 & 0.00 & 0.35 & 0.00 & \underline{0.36} & N/A \\
Qwen 3.5 Plus~\citep{qwen_35}     & 0.06 & 0.05 & 0.10 & 0.15 & 0.32 & 0.80 \\
\bottomrule
\end{tabular}
\vspace{-1.5em}
\end{table}

\vspace{-0.5em}
\subsection{Evaluation Result with CodegenBench}
\vspace{-0.5em}
The result is presented in Table~\ref{tab:result_x86_sunway} and ~\ref{tab:result_kunpeng}. LLMs demonstrate undeniable potential in high-performance code generation; however, this capability remains fundamentally constrained by the target architecture. In the x86 BLAS evaluation, the DeepSeek series generally met expectations. The latest DeepSeek-V4-Pro achieved the highest performance within the series, closely followed by the highly efficient DeepSeek-V4-Flash, while their predecessor, DeepSeek-V3.2, exhibited the weakest results. Concurrently, the Qwen models demonstrated competitive performance on par with the DeepSeek variants. Among closed-source contenders, the Claude series consistently delivered the best performance.

Nevertheless, this proficiency rapidly diminishes when shifting away from dominant architectures. The Kunpeng BLAS results clearly indicate a substantial performance degradation across all models as the target platform transitions from x86 to ARM. This challenge is further magnified in the LeetKunpeng and LeetSunway benchmarks, where all models struggled significantly with the $Pass@k$ metric, rendering most of the generated code unusable. The primary obstacle lies in the scarcity of training data: although extensive architecture documentation exists in the public domain for platforms like Kunpeng and Sunway, the critical lack of real-world open-source code examples severely hinders the models' ability to autonomously utilize unique hardware features. This directly reflects the exceptionally low $Pass_k$ rates observed in the Leet series benchmarks.

\subsection{Discussions and Analysis}

\begin{figure}[t]
  \centering
    \begin{minipage}{0.5\linewidth}
        \centering
        \includegraphics[width=\linewidth]{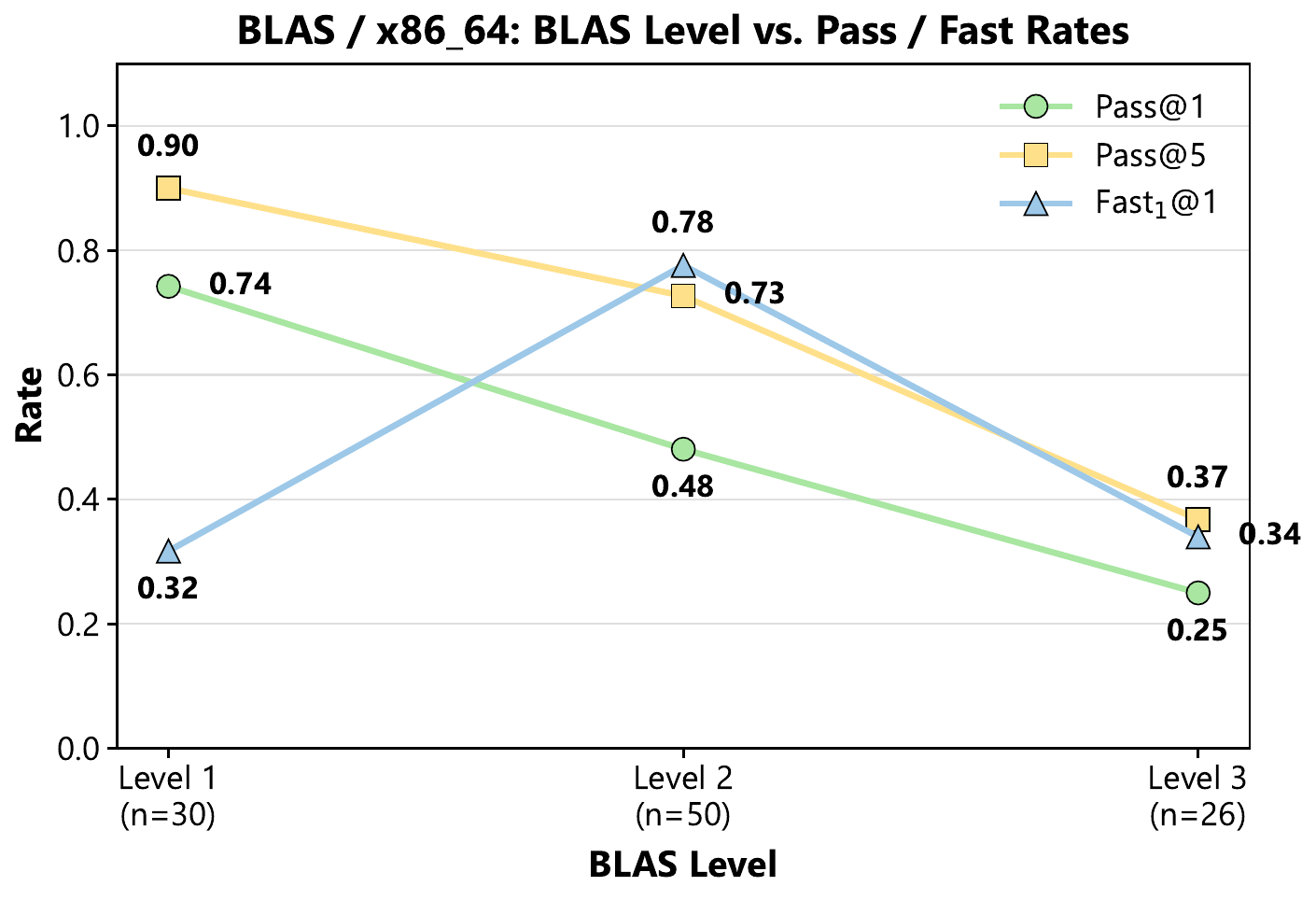}
        \label{fig:ltc}
    \end{minipage}
    \begin{minipage}{0.5\linewidth}
        \centering
        \includegraphics[width=\linewidth]{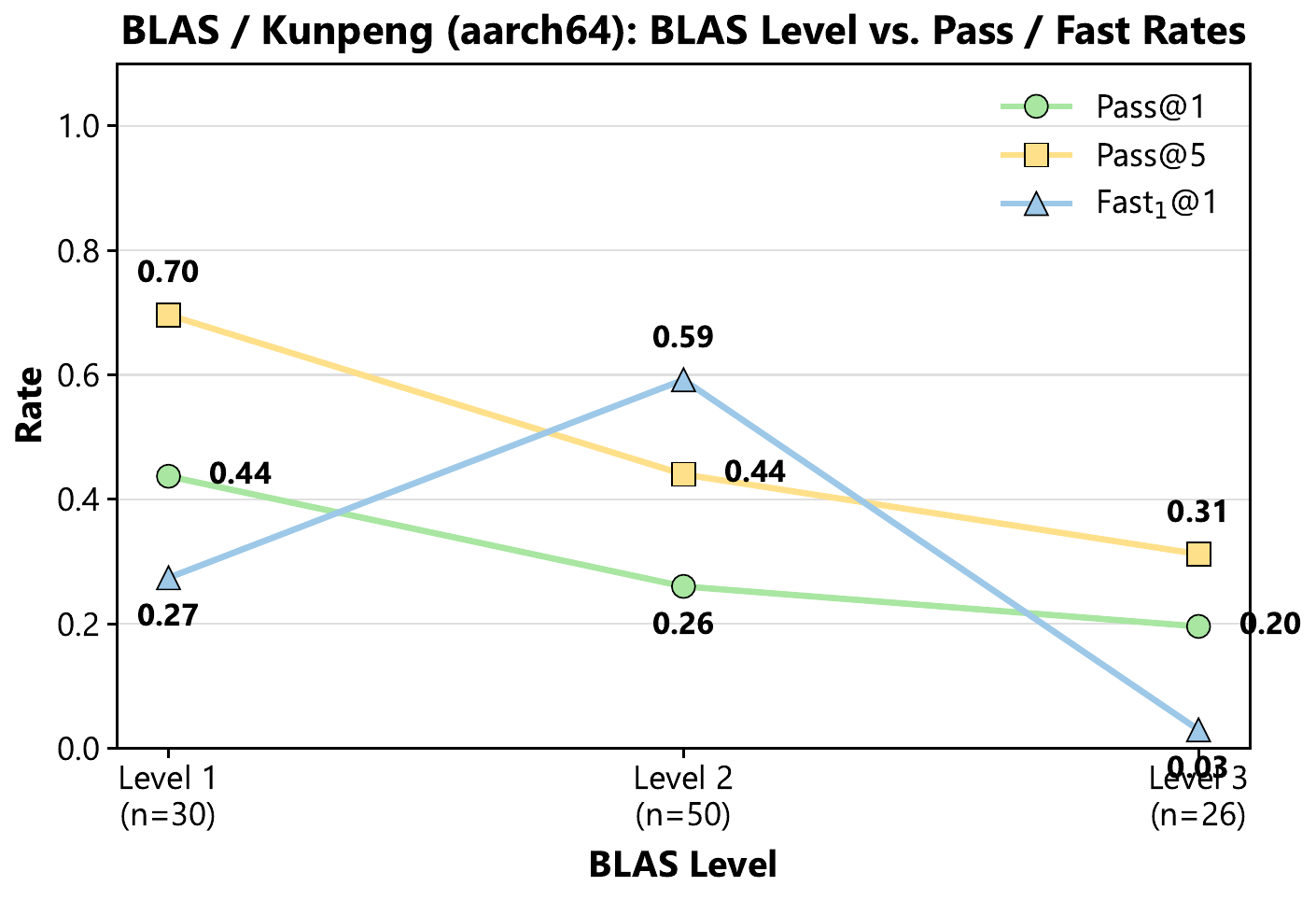}
        \label{fig:ltc_kp}
    \end{minipage}
    \vspace{-1.5em}
    \caption{Comparison of $Pass@1$, $Pass@5$, and $Fast_1@1$ metrics across varying levels of BLAS routines. Generally the complexity of BLAS routines increase as the Level increases.}
    \vspace{-1em}
    \label{fig:ltcs}
\end{figure}
\textbf{Influence of Task Complexity on Optimization} BLAS routines are formally divided into three complexity classifications: Level 1 (scalar and vector computations), Level 2 (matrix-vector operations), and Level 3 (matrix-matrix operations). Fig.~\ref{fig:ltcs} reveals an inverse correlation between task complexity and the $Pass@k$ metric, indicating that multi-dimensional data layouts introduce significant logical hurdles for models. Interestingly, the $Fast_1@1$ metric exhibits a divergent trend: LLM-generated code achieved comparatively better optimization speedups for Level 2 and Level 3 BLAS operations. This phenomenon likely stems from the fact that Level 1 routines are strictly memory-bandwidth bound with minimal structural complexity, leaving little room for models to introduce meaningful algorithmic optimization compared to the compute-bound nature of the higher-level routines.

\begin{figure}[t]
\centering
\includegraphics[width=\linewidth]{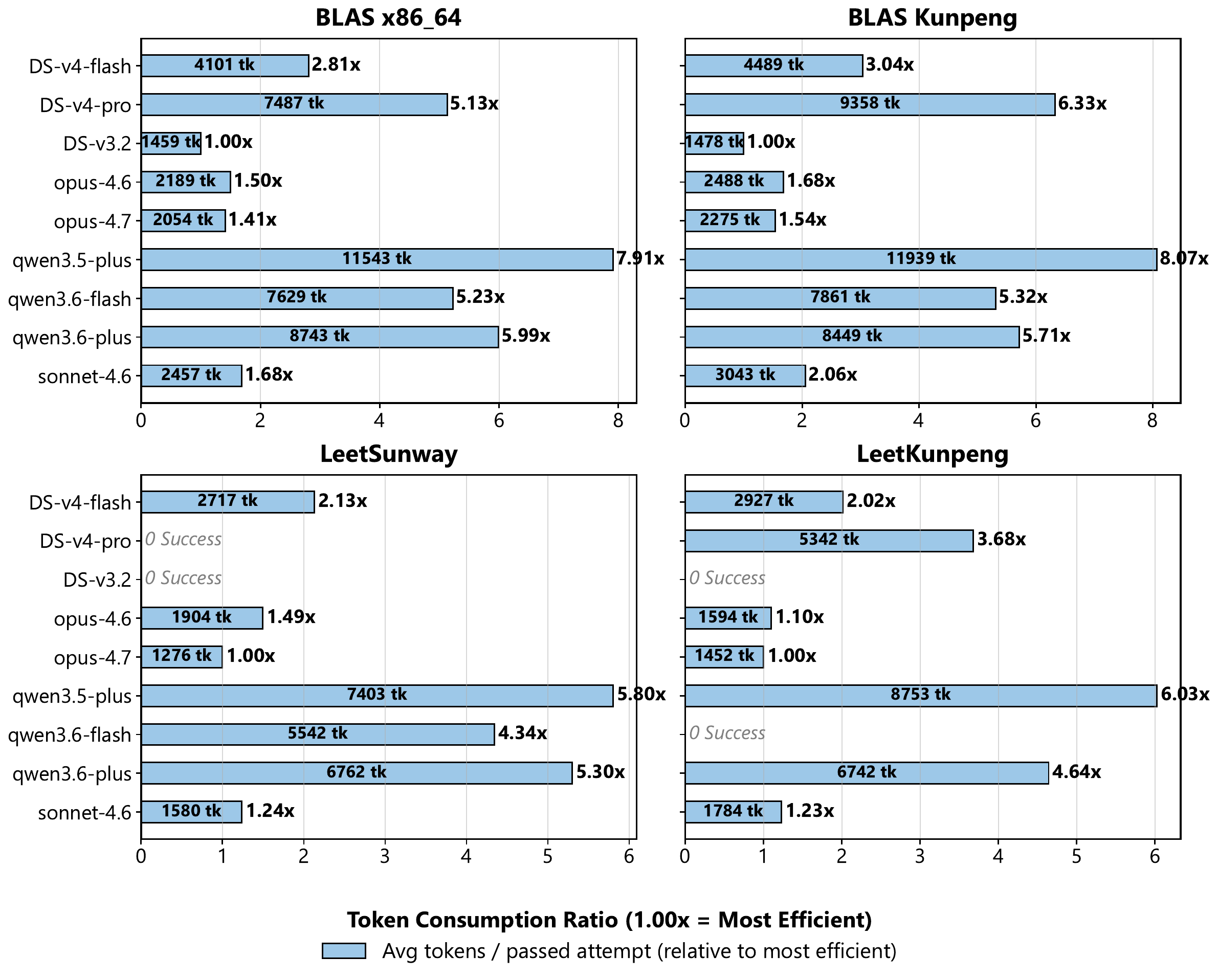}
\vspace{-2em}
\caption{Average tokens consumed to generate correct code. Claude series use fewer tokens to produce correct result, while their open-source contenders take more effort to achieve the same result.}
\label{tokenconsume}
\vspace{-1em}
\end{figure}
\textbf{Model Scale and Token Consumption vs. Performance.}
As illustrated in Fig.~\ref{tokenconsume}, the relationship between token consumption and performance gain is non-linear. Within the DeepSeek V4 family, the Pro variant, despite possessing a significantly larger parameter count, consumes nearly twice as many tokens as the Flash version, yet the corresponding performance improvement does not justify the additional computational overhead. While cross-model comparisons are influenced by various factors including model architecture, scale, and training objectives, these observations highlight a consistent trend among state-of-the-art LLMs: simply increasing model size or token expenditure does not guarantee proportional gains in code optimization capabilities.

\begin{figure}[t]
  \centering
    \includegraphics[width=\linewidth]{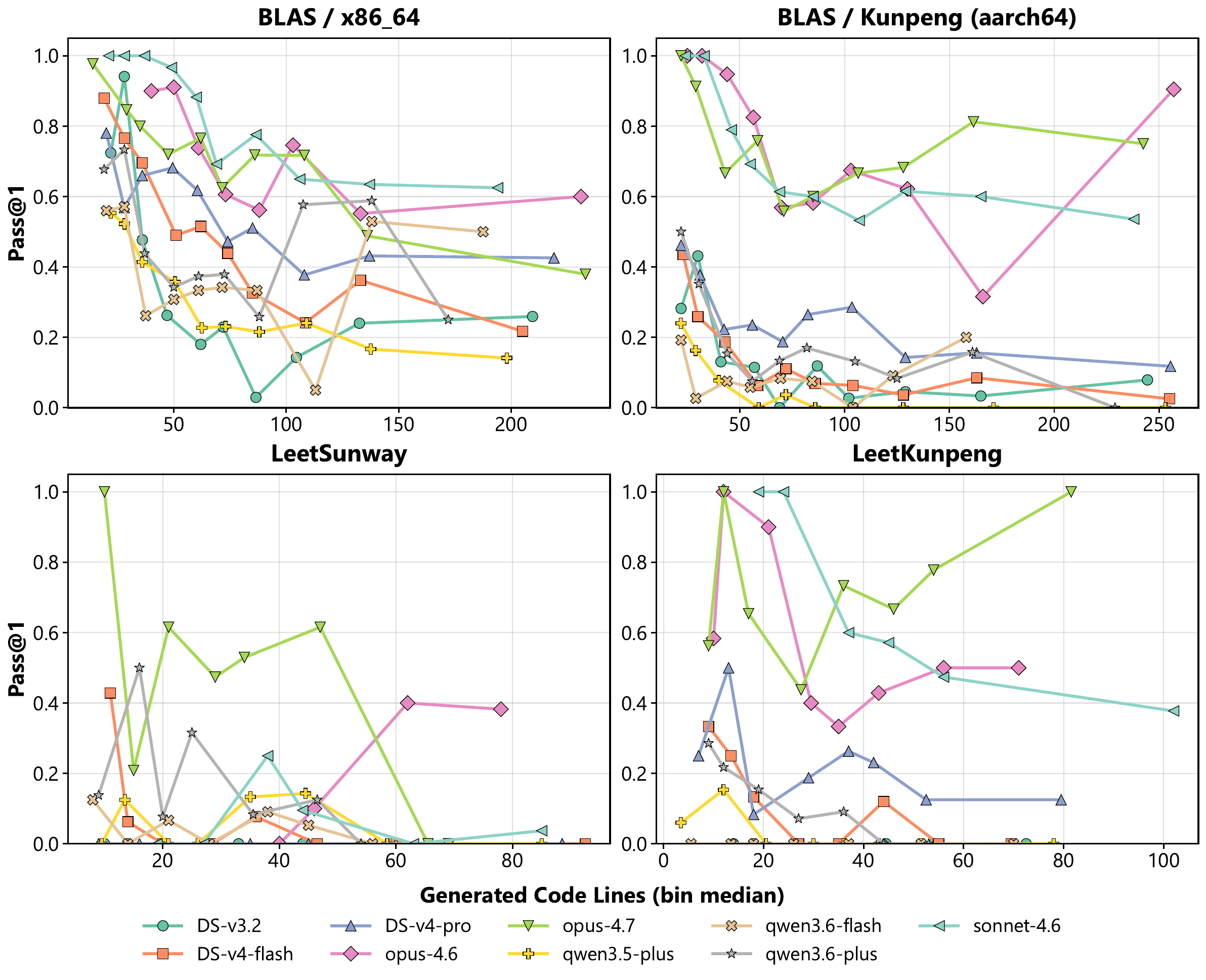}
    \vspace{-2em}
    \caption{Code Length vs. $Pass@1$, in four scenarios. $Pass@1$ decreases as code length increases. We also notice that there is a gap between open-source models and closed-source models in BLAS/Kunpeng configuration.}
    \label{fig:efforts}
    \vspace{-1.5em}
\end{figure}

\textbf{Code Generation Correctness vs. Code Length.} 
The $Pass@1$ metric assesses a model's intrinsic ability to generate contextually correct and executable code on its first attempt. Intuitively, as the length and complexity of the required implementation increase, the probability of encountering logical or syntactical errors grows. Fig.~\ref{fig:efforts} corroborates this expected trend. In the BLAS evaluations, all tested models demonstrated greater reliability when generating concise snippets (typically under 50 lines); however, passing rates degraded sharply as the code volume expanded. In the Leet series benchmarks, $Pass@1$ scores across all models remained uniformly low, likely exacerbated by the prolonged code length needed for manual architecture-specific implementations combined with a lack of open-access reference material.

\textbf{Evaluating Cross-Architecture Code Generation Efficacy.}
Let us recall our original question: \textit{Can LLMs genuinely write efficient code across diverse architectures? }
Based on our empirical findings, the answer is conditional. They demonstrate strong capabilities on ubiquitous platforms like x86, capitalizing on vast amounts of training data. Conversely, operating on specialized architectures with constrained open source representation remains a significant challenge. However, a compelling anomaly emerges within the evaluation metrics of the Leet benchmarks: while absolute generation correctness ($Pass@k$) remains poor, the $Fast_1@1$ score for successfully bounded generations ranks remarkably high. This is attributed to the Leet benchmark design, where LLMs are explicitly prompted to generate optimized implementations referencing basic baseline counterparts. As long as the LLM synthesizes semantically correct code that activates native hardware features, outperforming the unoptimized baseline CPU code proves relatively trivial. Conversely, in the BLAS tasks, model outputs face a much stricter threshold, as they are measured against highly hand-tuned standard libraries (e.g. OpenBLAS) employing intricate cache blocking and advanced vectorization techniques.
\vspace{-1.0em}
\section{Conclusion}
\vspace{-1.0em}
In this work, we introduced \textbf{CodegenBench}, a benchmark suite for evaluating LLM code generation across three hardware platforms: x86\_64, Kunpeng, and Sunway, with multi-architecture benchmark data comprising benchmark tasks that diverge from standard AI-centric workloads. We conduct an extensive empirical evaluation of 9 top-tier LLMs for CodegenBench. Our experiments reveal that the performance of LLMs is dependent on the target platform. Performance degradation is most severe on the Sunway architecture, which has the least public documentation and code examples, making automatic optimization and porting impractical for real-world use. In the future, we will continue to develop CodegenBench, working to improve LLMs' capabilities in generating high-quality efficient code for architectures like Sunway and Kunpeng.



\medskip

\small

\bibliographystyle{plainnat}
\bibliography{refs}


\newpage

\appendix

\begin{center}
    \Large
    \textbf{CodegenBench: Can LLMs Write Efficient Code \\Across Architectures? \\ (Supplemental Materials)}  
\end{center}

\startcontents[sections]
\printcontents[sections]{}{1}{\section*{Table of Contents in Appendix}\setcounter{tocdepth}{3}}

\newpage

\section{Limitations}
Despite the contributions of CodegenBench in establishing a multi-architecture benchmark for code efficiency using performance-oriented tasks, our study and benchmark have several limitations that warrant discussion. Firstly, while performing evaluations on Kunpeng and Sunway does provide insights in measuring LLMs' capabilities in generating optimized programs for different architecture, the access-barrier of these two hardware may lack of representative for architectures with more popularity like SPARC and PowerPC. The LeetKunpeng and LeetSunway suites cover a wide range of kernels, from linear algebra computation to machine learning optimizer. But the quantity of the test data is relatively insufficient, 20 kernels may be unable to fully represent the workload on these platforms. When generating the code, considering the overall size of the problem to be generated, we chose 5 as the sample size. This sample size may be too small to accurately give the probability value of the $Pass@k$ metric, which may weaken the final conclusion to some extent. Furthermore, the scope of evaluated models in this study remains relatively constrained. Although we selected representative state-of-the-art models, our evaluation currently omits other highly effective proprietary models such as OpenAI's GPT series and Google's Gemini family, as well as a broader spectrum of small-to-medium-sized open-weight models. Finally, the differences in code generation difficulty under different architectures make it difficult to quantify and standardize the final results, which may lead to the final results being difficult to understand intuitively.

\section{Broader Impacts}
\paragraph{Positive Societal Impacts:}The development and application of benchmarking tools such as CodegenBench help improve the efficiency of code generated by LLM across multiple platforms. This is expected to have positive social impacts, such as: optimizing kernel runtime for shorter times, thereby reducing energy consumption and improving the efficiency of existing programs; and promoting the adoption of high-efficiency architectures like ARM. The ARM architecture has not been widely adopted in HPC due to its immature ecosystem, and compared to mainstream platforms, it has significantly fewer available applications and infrastructure. The evolution of LLM in automatically writing kernels will greatly accelerate the development of applications and infrastructure. CodegenBench also provides the research community with a clearer understanding of LLMs' limitations regarding code efficiency when it comes to generating code for another architecture.

\paragraph{Negative Societal Impacts:}While improvements in LLMs capabilities will be beneficial to the society in various aspects, saving energy and time for a wide range of application, such progress might carry potential negative societal impacts as well. As LLMs become more adept at generating optimized code across various architectures, there is a risk that LLMs could be made to create malicious application, since the deep understanding into architectures also means that applications require securities may be easily cracked with the help of LLMs. Besides malicious purposes about binary integrity, it also raises concerns that malwares aiming to launch cyberattacks, requiring knowledge in utilizing heterogeneous devices' computation power. Addressing these concerns requires comprehensive supervision over purposes, responsible development and deployment for LLMs, and alignment to human's code of ethics.

\section{Evaluation Setup}

\subsection{Metrics}
\label{app:metric}
The metrics we use are defined as follows:
\begin{equation}
    Pass@1=\frac{\text{Number of tasks passing all test cases in one generation}}{\text{Total number of generation attempts}}
    \label{pass_1}
\end{equation}

\begin{equation}
    Pass@5=\frac{\text{Number of tasks passing all test cases within 5 generation attempts}}{\text{Total number of evaluation tasks}}
    \label{pass_5}
\end{equation}

\begin{equation}
    Fast_1@1=\frac{\text{Number of } Pass@1 \text{ valid tasks with speedup } > 1}{\text{Total number of } Pass@1 \text{ valid tasks}}
    \label{fast_1_1}
\end{equation}

\subsection{Evaluation Environments}
\label{app:evaluation_environments}
\subsubsection{x86-64}The x86-64 platform is equipped with the high-performance Xeon Platinum 8488C Processor. Benefiting from a mature and full-featured instruction set architecture that includes advanced vector extensions standard in modern enterprise servers, it serves as a highly representative baseline. This robust and ubiquitous environment makes it an ideal and reliable candidate for evaluating fundamental LLM code generation capabilities specifically targeting broadly adopted x86 systems. 

For BLAS-related experiments on this platform, we utilize \texttt{gcc} to compile and link against the \texttt{OpenBLAS} library.

\subsubsection{Sunway}
\begin{figure}
    \centering
    \includegraphics[width=\linewidth]{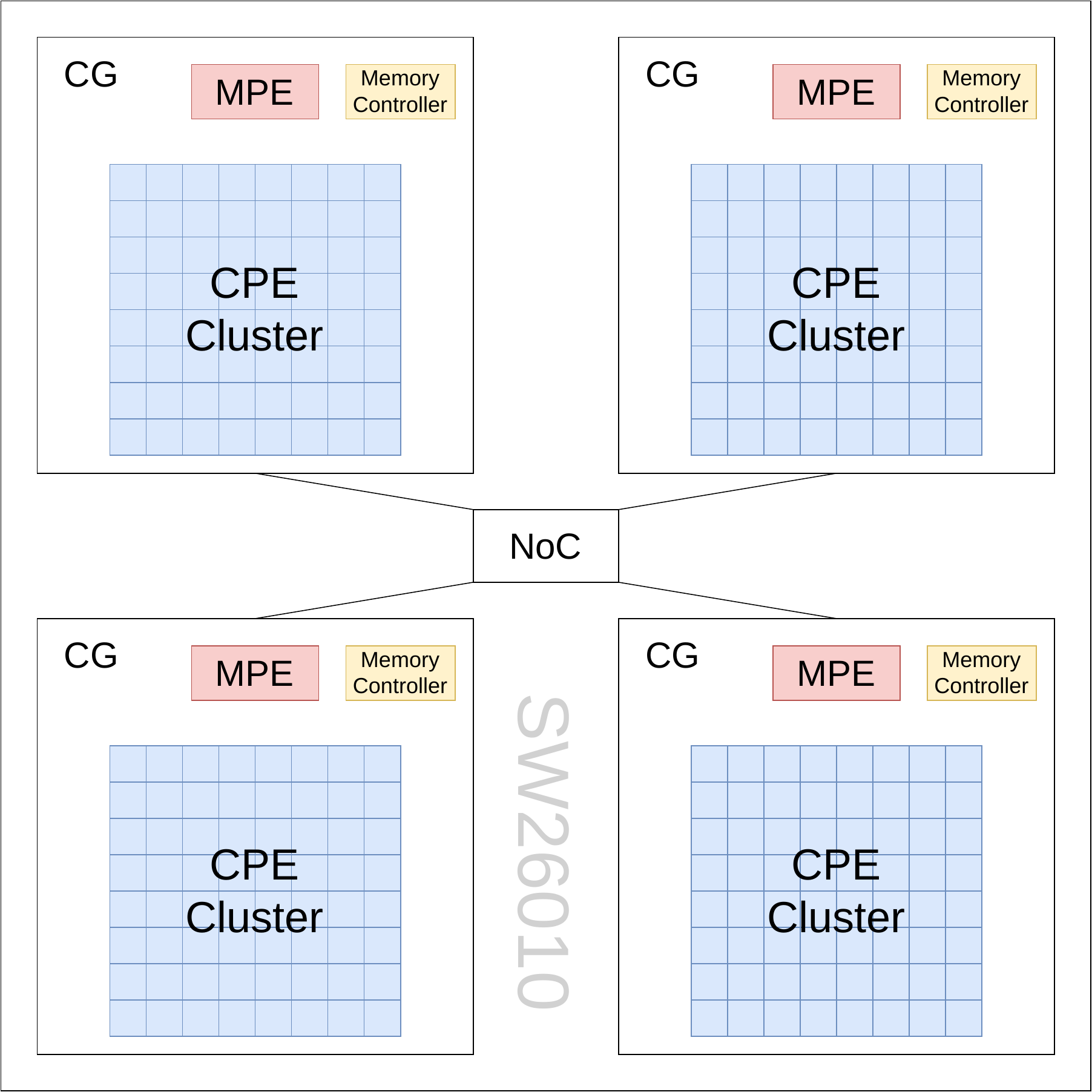}
    \caption{Architecture for SW26010.}
    \label{fig:sunway}
\end{figure}

The Sunway experiments are conducted on the renowned Sunway TaihuLight supercomputer infrastructure. This massive system is powered by 40,960 SW26010 many-core processors, complemented by a formidable overall aggregate memory capacity of 1,310,720 GB. As presented in Fig.~\ref{fig:sunway}, the processor's unique master-slave core paradigm and specialized memory hierarchy makes the Sunway architecture inherently well-suited for orchestrating massive, highly concurrent heterogeneous workloads. This presents a stark contrast to traditional x86 platforms, thereby offering a rigorous testbed for evaluating an LLM's capacity for parallel scheduling and architecture-specific code translation.

For all test cases on this architecture, all are compiled using the \texttt{swgcc} compiler to utilize the master-slave paradigm.

\subsubsection{Kunpeng}The Kunpeng platform, which utilizes the widely adopted ARM instruction set, serves as a highly competitive and increasingly prominent alternative to traditional x86-64 architectures in modern computing centers.

\begin{figure}
    \centering
    \includegraphics[width=\linewidth]{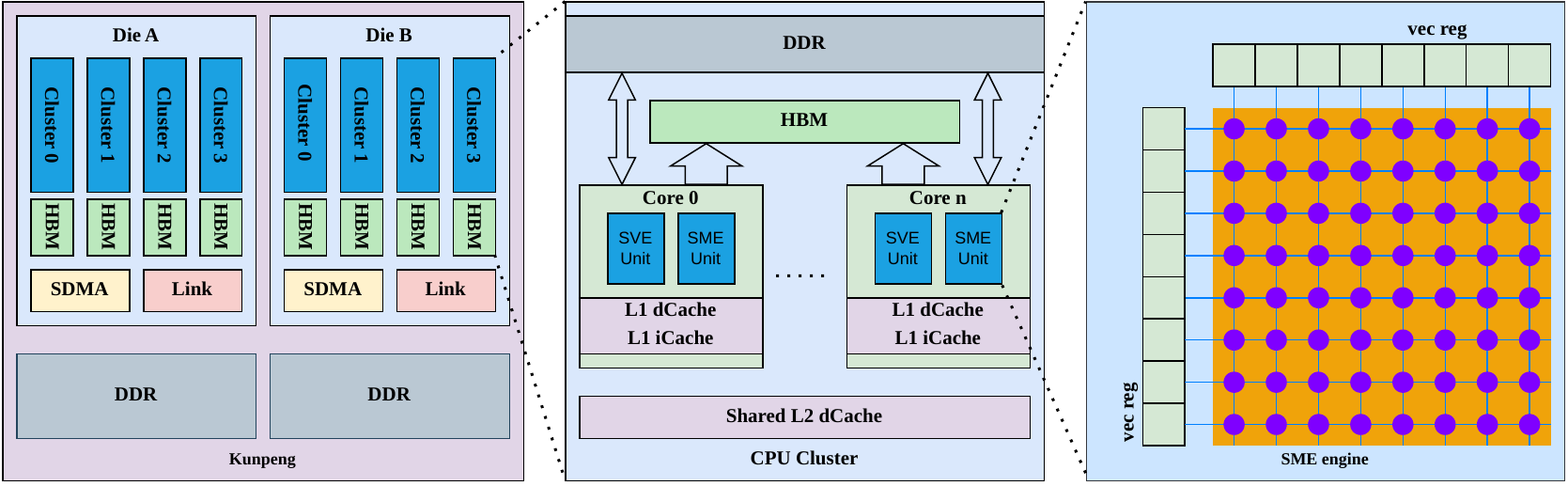}
    \caption{Architecture for Kunpeng.}
    \label{fig:kunpeng}
\end{figure}

Our experimental setup on this platform consists of a cluster of multiple interconnected nodes, where each individual node is robustly equipped with dual Kunpeng processors. Furthermore, as illustrated in Fig.~\ref{fig:kunpeng}, by featuring advanced parallel computing extensions such as SVE/SVE2 and SME, this ARM-based processor harbors immense performance potential, providing an excellent environment to assess how well LLMs can adapt to and exploit cutting-edge vector and matrix acceleration instructions. Besides, on-die HBM provides potential in creating programs solving issue in memory-bound bottleneck.

For BLAS evaluations on the Kunpeng architecture, we employ \texttt{bisheng}, a customized \texttt{clang} compiler, and link against the \texttt{kblas} library. LeetKunpeng evaluations are with \texttt{clang} compiler, utilizing the specialized on-die HBM feature in Kunpeng chip.

\subsection{Evaluation Models}
\label{app:evaluation_models}
\begin{table}[htbp]
\centering
\caption{Evaluated Models and Their References}
\label{tab:models_links}
\begin{tabular}{ll}
\toprule
\textbf{Model Name} & \textbf{Reference / Link} \\
\midrule
\multicolumn{2}{c}{\textit{Open-Source Models}} \\
\midrule
DeepSeek V3.2 & \url{https://huggingface.co/deepseek-ai/DeepSeek-V3.2} \\
DeepSeek V4 Flash & \url{https://huggingface.co/deepseek-ai/DeepSeek-V4-Flash} \\
DeepSeek V4 Pro & \url{https://huggingface.co/deepseek-ai/DeepSeek-V4-Pro} \\
Qwen 3.5 Plus & \url{https://qwen.ai/blog?id=qwen3.5} \\
Qwen 3.6 Flash & \url{https://qwen.ai/blog?id=qwen3.6-35b-a3b} \\
\midrule
\multicolumn{2}{c}{\textit{Closed-source Models}} \\
\midrule
Claude Sonnet 4.6 & \url{https://anthropic.com/claude/sonnet} \\
Claude Opus 4.6 & \url{https://anthropic.com/claude/opus} \\
Claude Opus 4.7 & \url{https://anthropic.com/claude/opus} \\
Qwen 3.6 Plus & \url{https://qwen.ai/blog?id=qwen3.6} \\
\bottomrule
\end{tabular}
\end{table}

As listed in Sec.~\ref{sec:exp_setup}, we evaluate both closed-source and open-source models that represent the current state-of-the-art in code generation. Our selection rationale is grounded in the distinct strengths of these model families. The Claude series is widely recognized as one of the most capable specialized programming models available, serving as the foundational engine for advanced agentic frameworks like Claude Code. Among open-source offerings, the DeepSeek series has distinguished itself by achieving performance on par with leading closed-source models at a highly efficient cost; notably, its latest iteration, DeepSeek V4 Pro, yields competitive results on rigorous benchmarks such as SWE-Bench\cite{jimenez2023swe}, rivaling even Claude Opus 4.6. Finally, we incorporate the Qwen series, which provides a diverse spectrum of models across various parameter scales, establishing a comprehensive and robust baseline for our comparative analysis.

\paragraph{Closed-source Models.}We consider four closed-source models accessed through official APIs: Claude Sonnet 4.6, Claude Opus 4.6, Claude Opus 4.7 from Anthropic, Qwen 3.6 Plus from Qwen. For each model, we use the chat completion interface, and we supply a short system instruction that asks the model to answer in code with proper optimizing mechanism applied. We do not apply task-specific fine-tuning, so the reported results reflect the intrinsic zero-shot capabilities of these closed-source models on CodegenBench.

\paragraph{Open-source Models.}We select five representative models accessed through official APIs: DeepSeek V3.2, DeepSeek V4 Flash and DeepSeek V4 Pro from DeepSeek, Qwen 3.6 Flash and Qwen 3.5 Plus from Qwen. For each model, we apply same setting on propriety models, with the same system prompt. We do not apply task-specific fine-tuning, so the reported results reflect the intrinsic zero-shot capabilities of these open-source models on CodegenBench.

\section{Results for BLAS/x86\_64 and BLAS/Kunpeng}

\subsection{Performance Degradation in Complex Number Arithmetic}

\begin{figure}[t]
  \centering
    \begin{minipage}{0.5\linewidth}
        \centering
        \includegraphics[width=\linewidth]{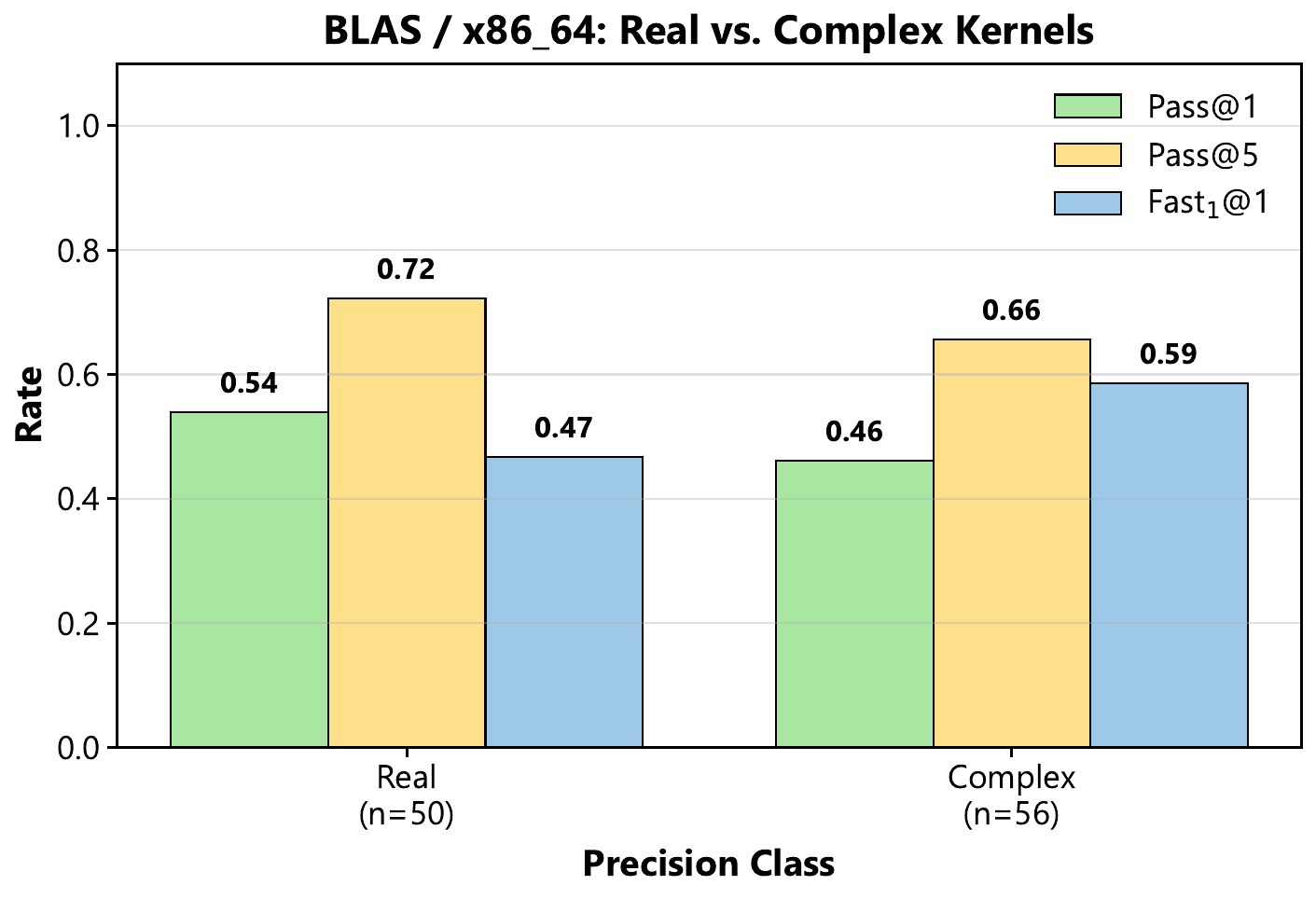}
        \label{fig:avg_rvc}
    \end{minipage}
    \begin{minipage}{0.5\linewidth}
        \centering
        \includegraphics[width=\linewidth]{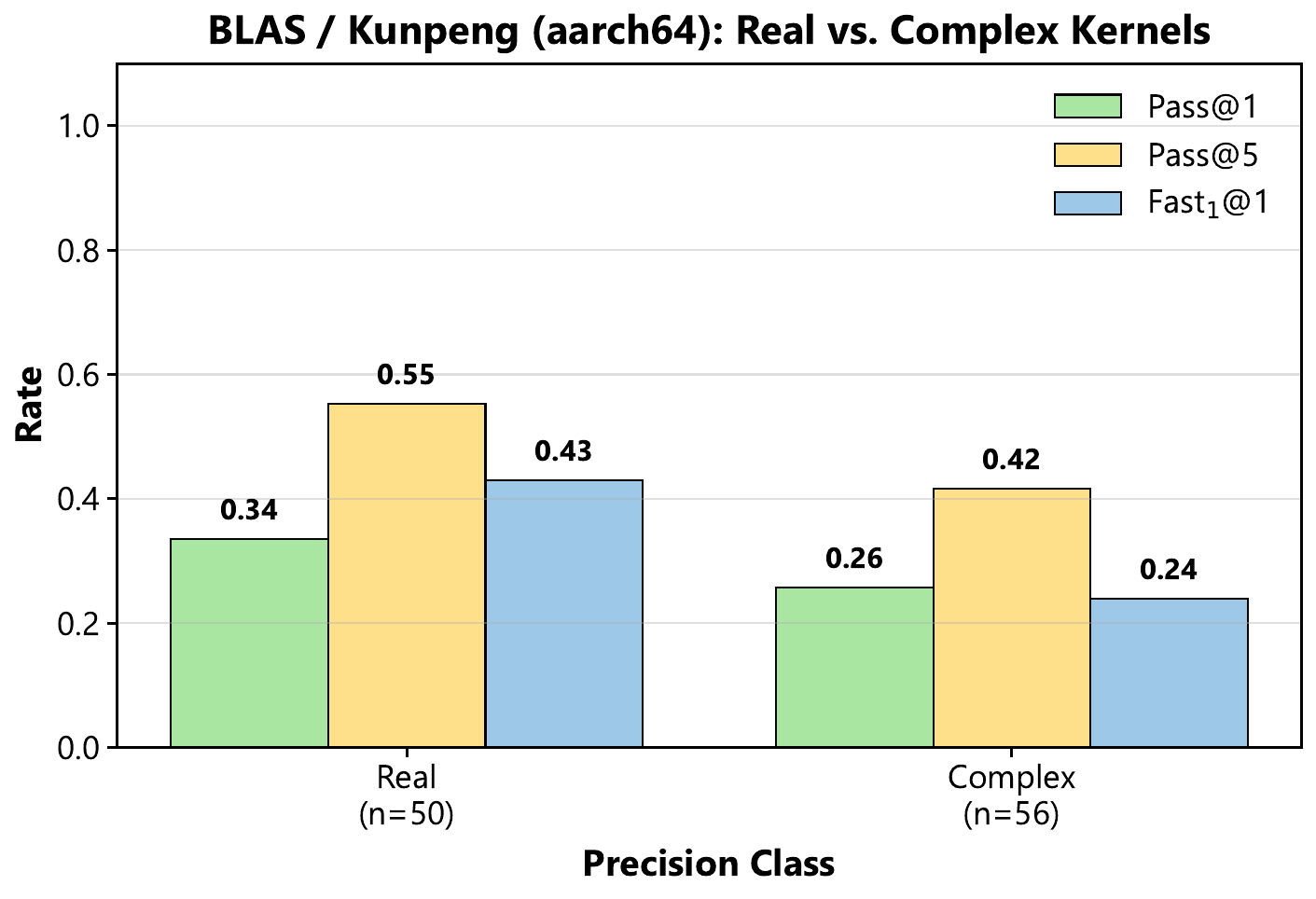}
        \label{fig:avg_rvc_kp}
    \end{minipage}
    \caption{Performance comparison distinguishing real and complex number arithmetic on BLAS/Kunpeng per model. While real arithmetic consistently maintains higher correctness ($Pass@1$), the execution acceleration metric ($Fast_1@1$) reveals distinct, architecture-dependent outcomes.}
    \label{fig:avg_rvcs}
\end{figure}

BLAS routines are fundamentally categorized into two computational domains: real number arithmetic and complex number arithmetic. Within the BLAS/x86\_64 evaluation environment, real-valued operations systematically outperform their complex-valued counterparts in the $Pass@1$ metric across all evaluated models as presented in Fig.~\ref{fig:avg_rvcs}, registering an average advantage of approximately 8 percentage points. Interestingly, this hierarchy is entirely inverted when evaluating the $Fast_1@1$ metric; every model demonstrates superior acceleration capabilities on complex arithmetic tasks compared to real ones. This phenomenon indicates that the reference implementations for complex routines likely possess lower baseline execution efficiencies, thereby presenting a more accessible threshold for LLM-driven performance improvements. 

\subsection{Model-Level Performance Degradation On x86\_64}
\begin{figure}[t]
  \centering
    \includegraphics[width=\linewidth]{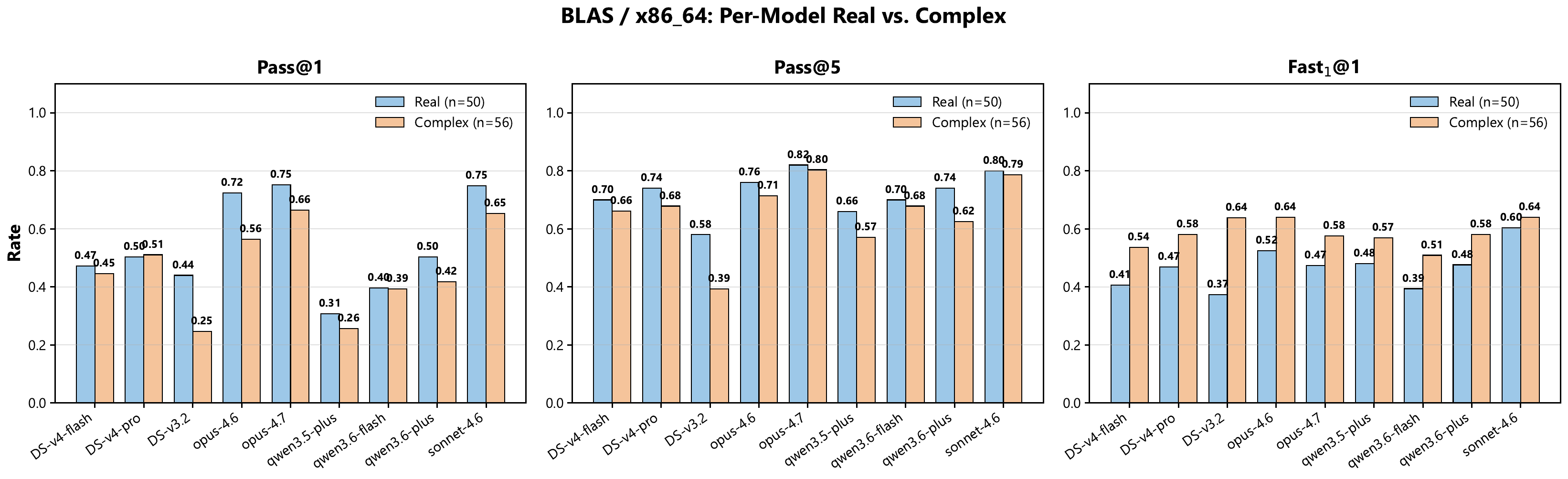}
    \caption{Performance comparison distinguishing real and complex number arithmetic on BLAS/x86\_64 per model. While real arithmetic consistently maintains higher correctness ($Pass@1$), the execution acceleration metric ($Fast_1@1$) reveals distinct, architecture-dependent outcomes.}
    \label{fig:rvcp}
\end{figure}
Detailed model-level analysis within the BLAS/x86\_64 environment corroborates the aggregate trends regarding arithmetic domains. As illustrated in Fig.~\ref{fig:rvcp}, $Pass@1$ for real-valued arithmetic consistently surpasses that of complex-valued operations across the entire model roster. Conversely, $Fast_1@1$ exhibits a pronounced inverse trend: code generated for complex arithmetic yields systematically higher speedup ratios. This divergence underscores a distinct characteristic of LLM code generation on x86\_64 architectures: while complex mathematical logic introduces structural hurdles leading to compilation or validation failures, successfully generated complex routines benefit significantly from mature, LLM-driven vectorization and memory optimizations, allowing them to achieve impressive execution accelerations even when rigorously benchmarked against highly optimized libraries like \texttt{OpenBLAS}.

\subsection{Model-Level Performance Degradation On Kunpeng}
\begin{figure}[t]
  \centering
    \includegraphics[width=\linewidth]{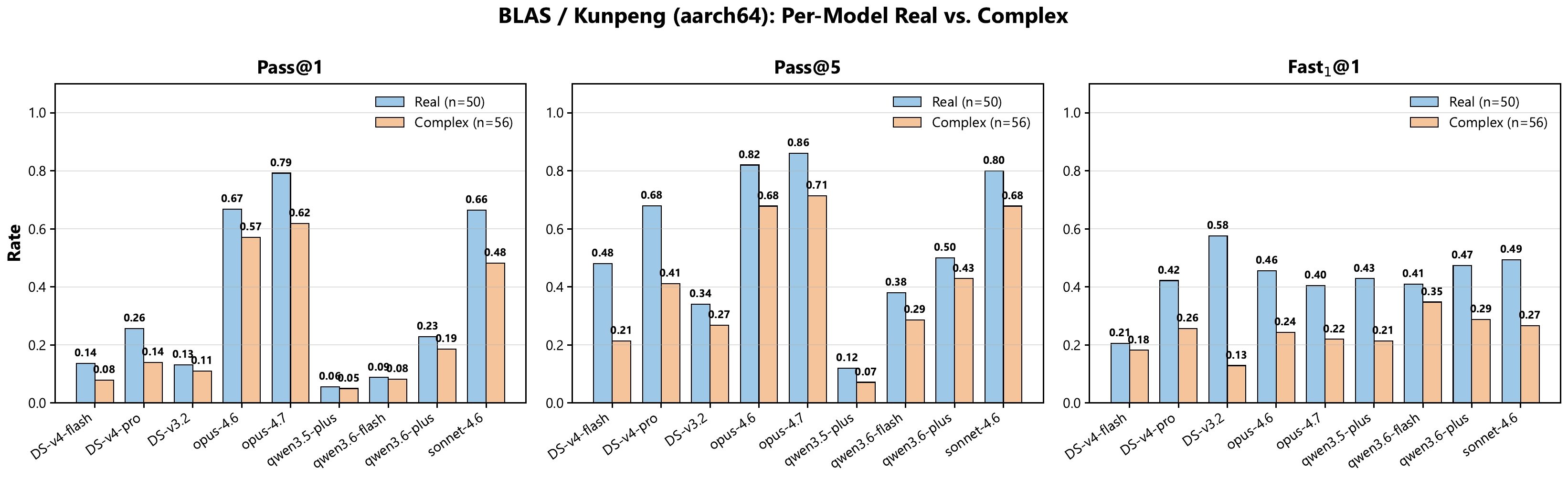}
    \caption{Performance comparison distinguishing real and complex number arithmetic across BLAS/x86\_64 and BLAS/Kunpeng evaluation environments. While real arithmetic consistently maintains higher correctness ($Pass@1$), the execution acceleration metric ($Fast_1@1$) reveals distinct, architecture-dependent outcomes.}
    \label{fig:rvcpkp}
\end{figure}
Conversely, within the BLAS/Kunpeng case, while the identical correctness pattern persists, with real arithmetic consistently surpassing complex arithmetic, and the $Fast_1@1$ dynamics diverge significantly from the x86\_64 results as presented in Fig.~\ref{fig:rvcpkp}. In this setting, real arithmetic generally dominates the acceleration metrics across the majority of models. DeepSeek v3.2 represents the sole exception, where complex arithmetic incidentally achieves a superior $Fast_1@1$ rate. Ultimately, the systemic deficiency in leveraging aarch64 SVE intrinsics for complex arithmetic is most conspicuously manifested in the substantial degradation of the $Pass@1$ and $Pass@5$ correctness scores.

\section{Case Studies}
\subsection{Successful Examples}
\begin{figure}[t]
\includegraphics[width=\linewidth]{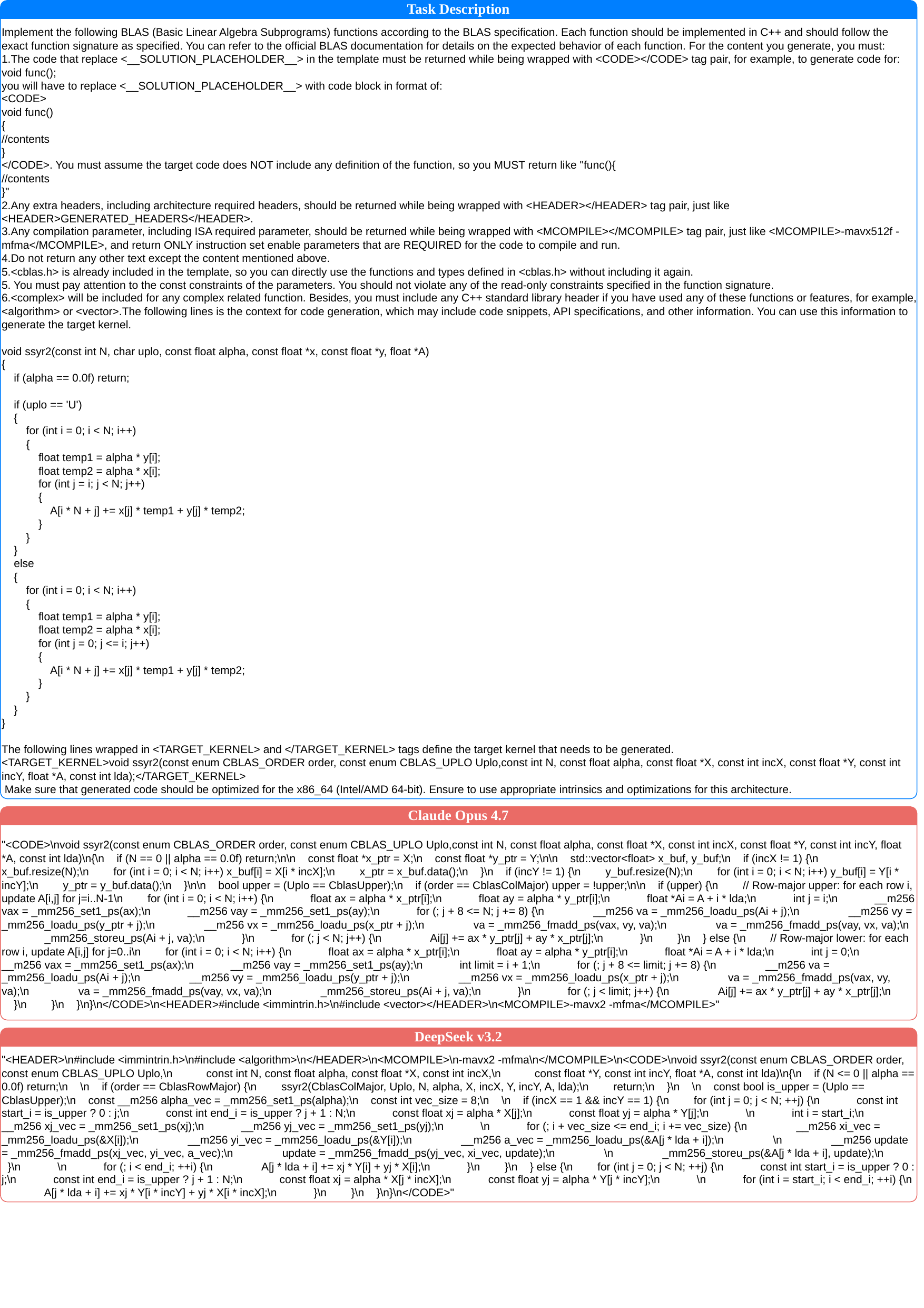}
\caption{\texttt{ssyr2} case in BLAS/x86, Both Opus and DeepSeek chose appropriate intrinsics and surpassed \texttt{OpenBLAS}.}
\label{case_ssyr2}
\end{figure}

\paragraph{Successful Example 1:} As illustrated in Fig.~\ref{case_ssyr2}, we present a case study evaluating the \texttt{ssyr2} operation within the BLAS/x86 environment. In this scenario, the solutions synthesized by both Claude Opus 4.7 and DeepSeek v3.2 significantly outperform the highly-optimized \texttt{OpenBLAS} reference implementation. 

Both models predominantly selected the AVX2 instruction set, effectively capitalizing on the advanced vectorized arithmetic capabilities inherent to modern x86\_64 processors. Specifically, Opus and DeepSeek converged on utilizing the \texttt{\_mm256\_fmadd\_ps} fused multiply-add intrinsic, demonstrating a consistent and highly efficient architectural pattern for this workload.

Achieving this level of performance necessitates meticulous management of strided memory accesses (\texttt{incx} and \texttt{incy}) alongside the judicious application of advanced vector instruction sets, such as AVX, tailored to the target architecture. This empirical evidence underscores the capacity of contemporary LLMs to not only navigate intricate edge cases but also to architect computationally superior, high-fidelity algorithmic solutions.

\begin{figure}[t]
\includegraphics[width=\linewidth]{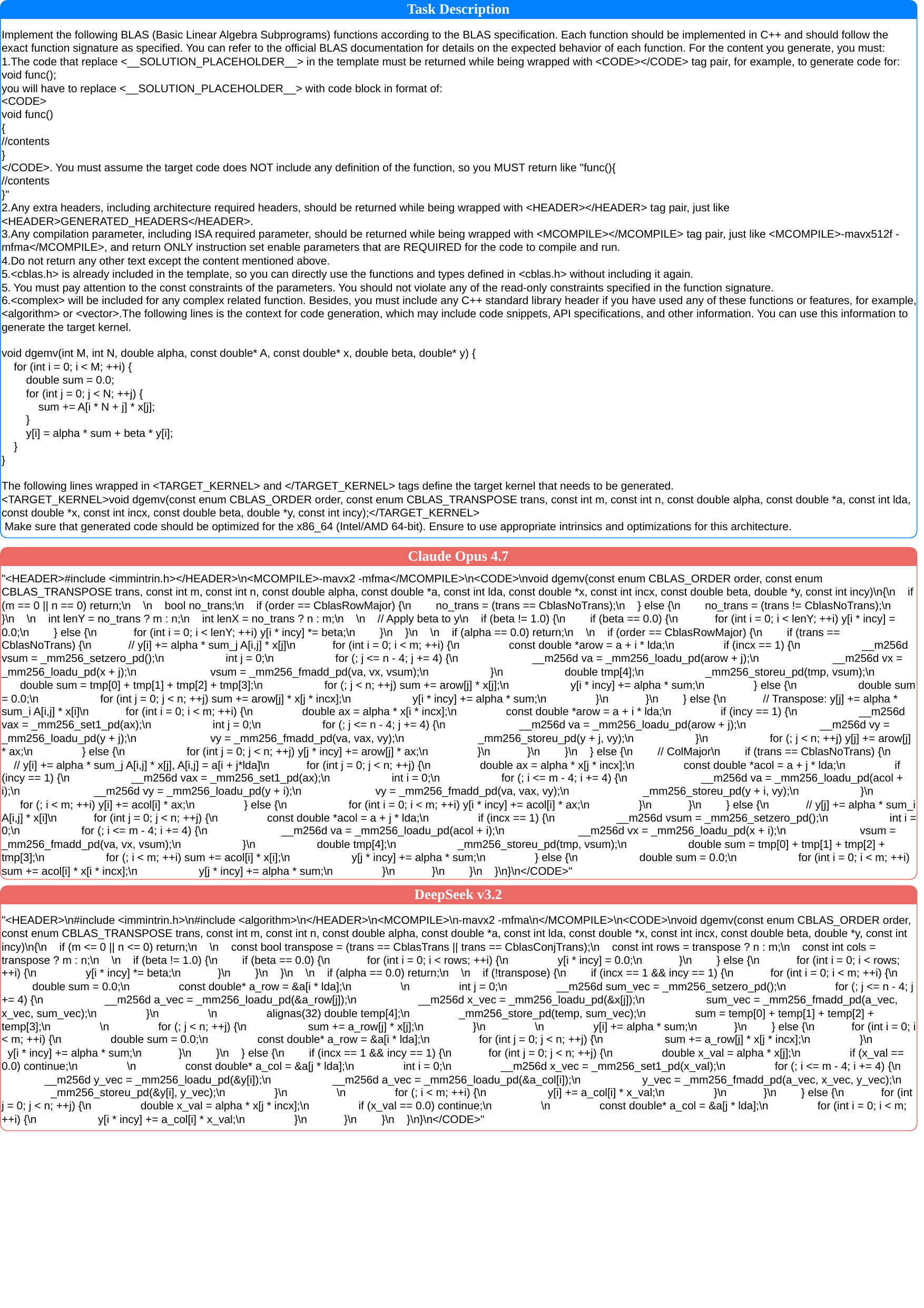}
\caption{\texttt{dgemv} case in BLAS/x86, Both Opus and DeepSeek chose appropriate intrinsics and surpassed \texttt{OpenBLAS}.}
\label{case_dgemv}
\end{figure}

\paragraph{Successful Example 2:} Fig.~\ref{case_dgemv} demonstrates another compelling instance of LLM-driven optimization for the \texttt{dgemv} routine under the BLAS/x86 setting. Similar to the previous case, code generated by Claude Opus 4.7 and DeepSeek v3.2 exhibits execution efficiency superior to that of \texttt{OpenBLAS}. 

When tackling the \texttt{dgemv} benchmark, both Opus and DeepSeek similarly converged on deploying the \texttt{\_mm256\_fmadd\_ps} intrinsic, mirroring their strategy from the \texttt{ssyr2} evaluation. This outcome strongly implies that for linear algebra kernels characterized by regular, predictable memory access patterns, state-of-the-art LLMs can seamlessly pair algorithmic logic with the optimal vectorized intrinsics available on x86\_64 platforms.

This specific task imposes stringent requirements on the accurate orchestration of multi-dimensional matrix-vector multiplications, precise stride (\texttt{inc}) computations, and the optimal employment of AVX vectorized routines. The results further validate the proficiency of LLMs in producing extremely performant code while strictly adhering to complex operational constraints.

\begin{figure}[t]
\includegraphics[width=\linewidth]{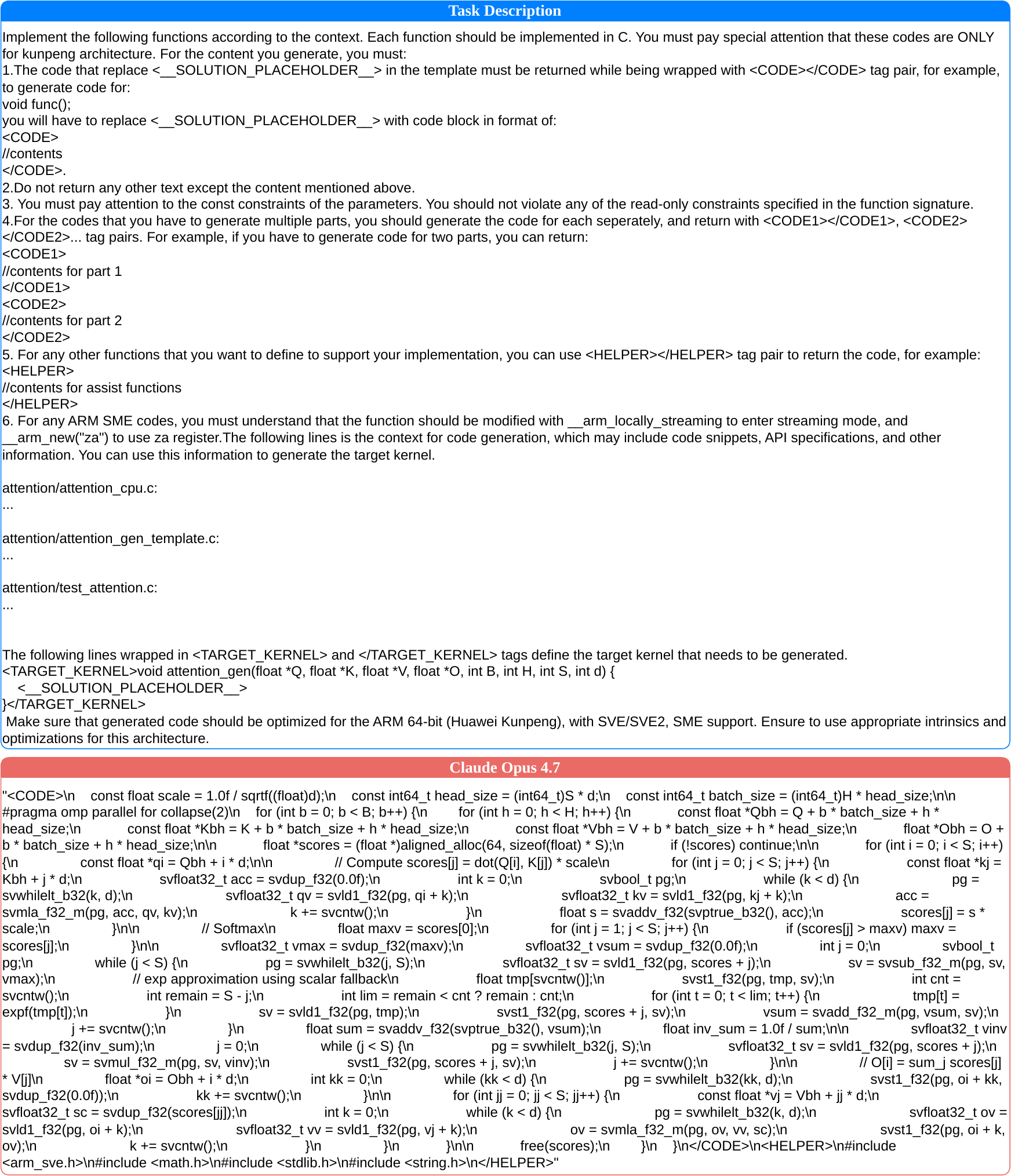}
\caption{\texttt{attention} case in LeetKunpeng, Opus chose appropriate intrinsics and surpassed referenced implementation.}
\label{case_attention}
\end{figure}

\paragraph{Successful Example 3:} Fig.~\ref{case_attention} demonstrates attention kernel optimization on Kunpeng. Claude Opus 4.7 chose proper SVE intrinsics to accelerate the process of attention calculation.

The standard attention kernel is a memory-bound kernel, but it still leaves some room for vectorization optimization. Claude Opus 4.7 achieves a level of $Pass@1=1$ in attention kernel generation, demonstrating its powerful ability to generate typical kernel functions suitable for modern LLM architectures. While generating correct kernels, these optimized code achieved at least $1.6\times$ speedup in FLOPS.

\subsection{Suboptimal Examples}
\begin{figure}[t]
\includegraphics[width=\linewidth]{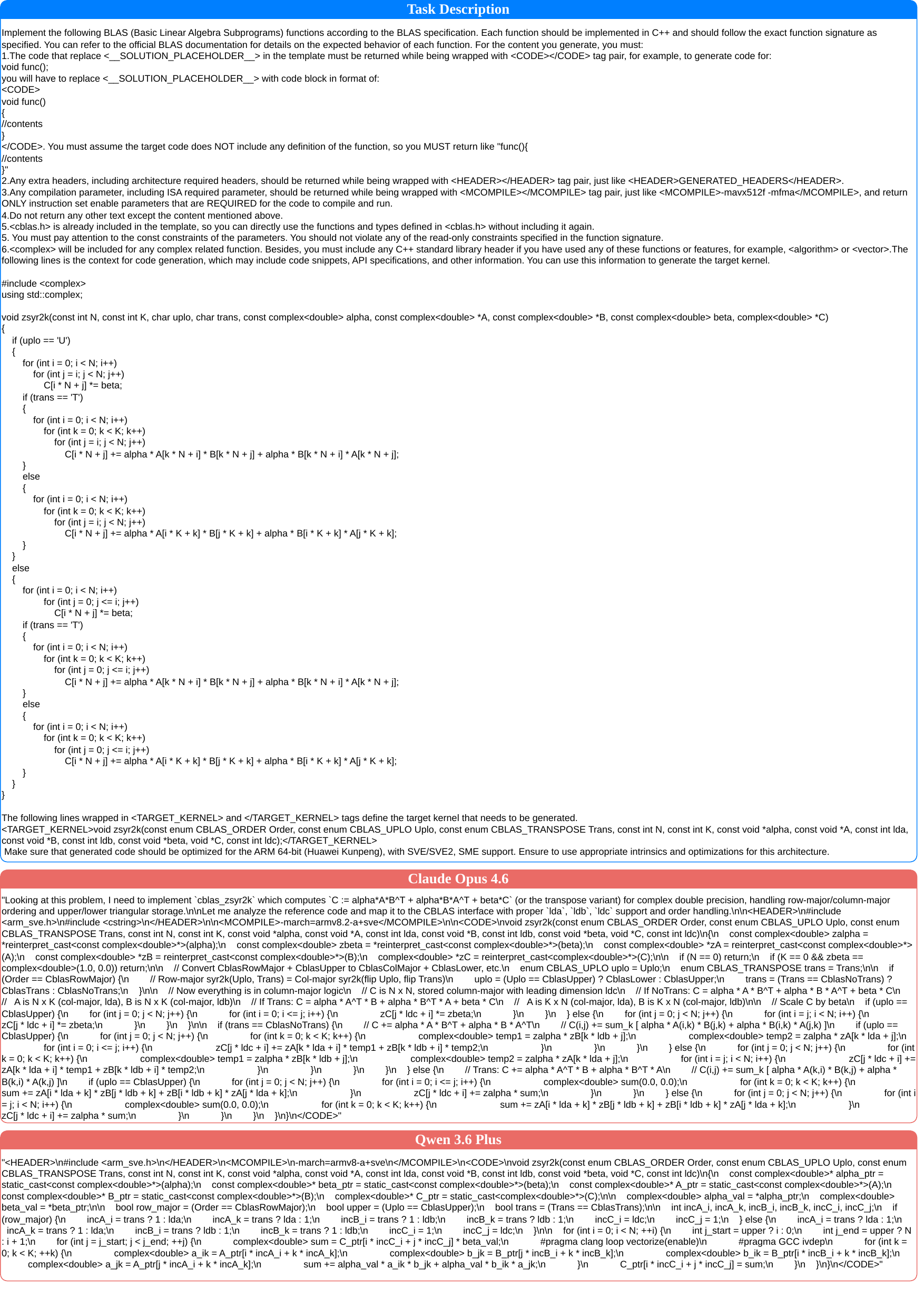}
\caption{\texttt{zsyr2k} case in BLAS/Kunpeng, Opus and Qwen provided correct result, but the execution efficiency were much worse than kblas under BLAS/Kunpeng setting.}
\label{case_zsyr2k}
\end{figure}

\paragraph{Suboptimal Example 1:} Conversely, Fig.~\ref{case_zsyr2k} details a scenario evaluating the \texttt{zsyr2k} operation where LLM-generated code falls short of expected efficiency benchmarks. While the implementations provided by Claude Opus 4.6 and Qwen 3.6 Plus strictly adhere to mathematical correctness, a non-trivial feat given the inherent complexities of floating-point complex number arithmetic, their runtime performance is drastically suboptimal.

It is crucial to note that contemporary instruction set architectures generally lack native, single-instruction support for complex number computational workloads. Consequently, LLMs face the structural burden of manually decomposing complex operations into corresponding real and imaginary scalar or vector computations. This extra layer of abstraction significantly complicates the generation of code that is concurrently accurate and hardware-efficient.

Specifically, within the BLAS/Kunpeng environment, the execution time of the code generated by Qwen was observed to be an order of magnitude (at least $10\times$) slower than the highly-tuned \texttt{kblas} library.

\begin{figure}[t]
\includegraphics[width=\linewidth]{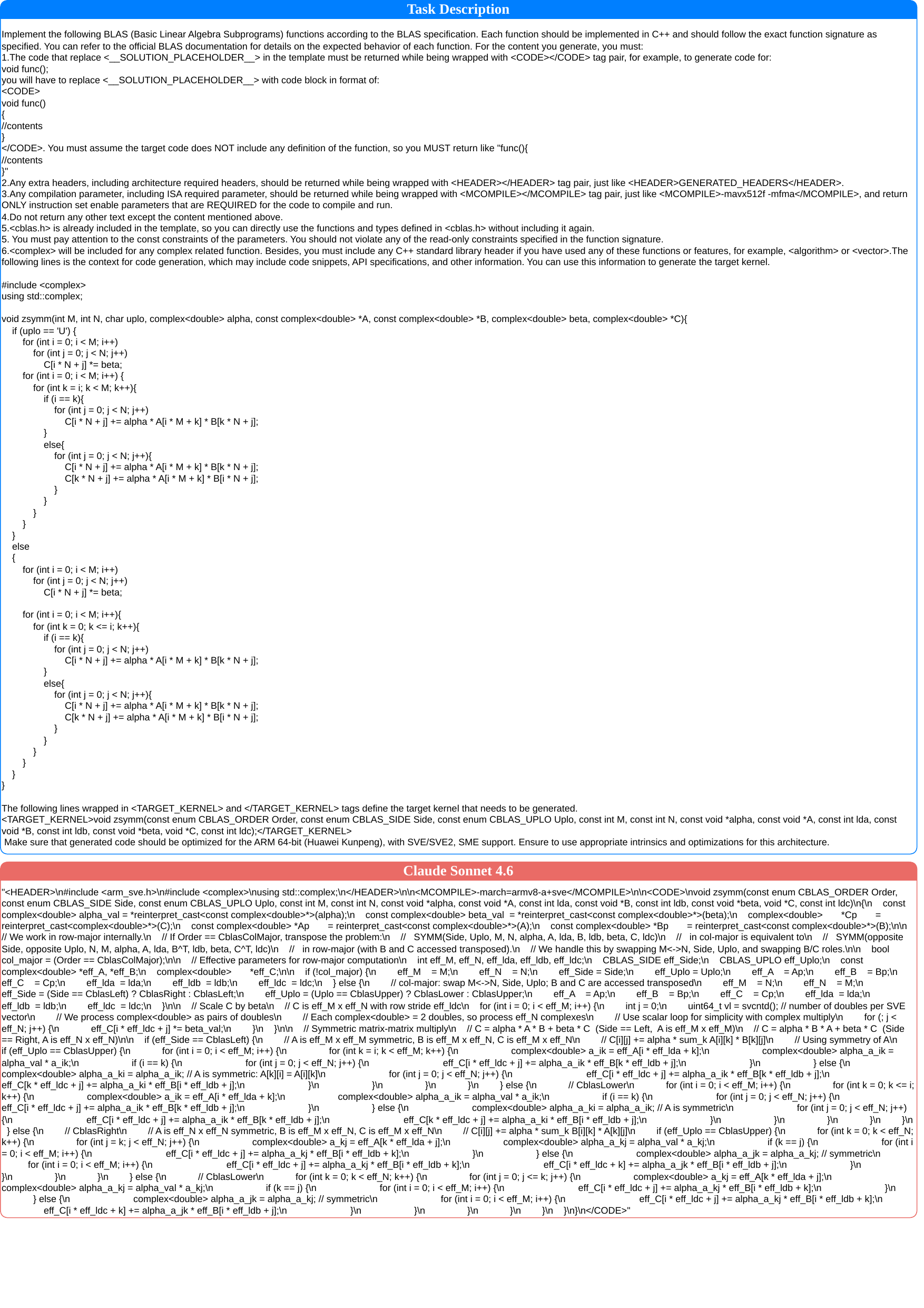}
\caption{\texttt{zsymm} case in BLAS/Kunpeng, Sonnet provided correct result, but the execution efficiency were worse than kblas under BLAS/Kunpeng setting.}
\label{case_zsymm}
\end{figure}

\paragraph{Suboptimal Example 2:} Another suboptimal case is observed in Fig.~\ref{case_zsymm}, which examines the \texttt{zsymm} function. Here, Claude Sonnet 4.6 successfully synthesizes a functionally accurate solution that robustly handles the required complex number arithmetic components. However, despite passing rigorous correctness validations, the execution latency remains a critical bottleneck. 

The fundamental challenge of \texttt{zsymm} involves managing scattered memory access patterns dictated by the symmetric matrix structure alongside the computational intensity of complex arithmetic. While established high-performance computing libraries like \texttt{kblas} employ sophisticated cache blocking and register tiling maneuvers dynamically tuned for complex data types, the LLM-generated code predominantly defaults to structurally naive, unoptimized loop traversals. Consequently, the synthesized implementation suffers from severe bottlenecks, executing more than $10\times$ slower when benchmarked against the \texttt{kblas} library.

\begin{figure}[t]
\includegraphics[width=\linewidth]{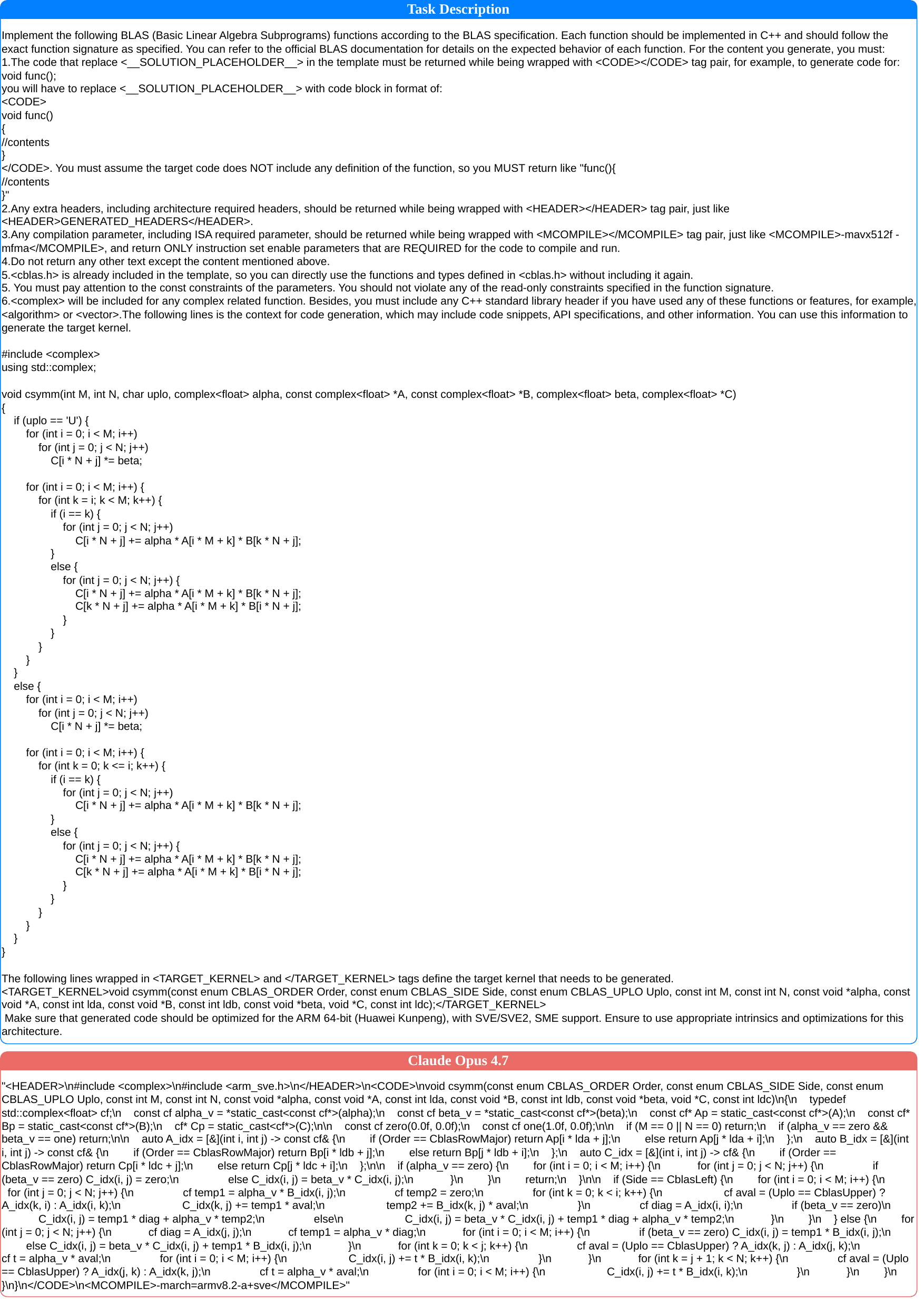}
\caption{\texttt{csymm} case in BLAS/Kunpeng, Opus provided correct result, but the execution efficiency were worse than kblas under BLAS/Kunpeng setting.}
\label{case_csymm}
\end{figure}

\paragraph{Suboptimal Example 3:} Building upon the previous observation, Fig.~\ref{case_csymm} illustrates a related scenario evaluating the \texttt{csymm} function. In this evaluation, Claude Opus 4.7 similarly synthesizes a mathematically exact solution capable of robustly managing the requisite complex algebraic expressions. Nevertheless, obtaining acceptable execution latency persists as a structural challenge. 

Performance profiling indicates a behavioral execution pattern structurally parallel to the \texttt{zsymm} evaluation. The Opus model skillfully formulates the algebraic correctness but fundamentally fails to instantiate memory-aware vectorization templates for single-precision complex matrices. Deprived of these essential, hardware-aware micro-optimizations, the generated codebase coerces the processor to execute scalarized, highly sequential memory fetch operations. This forces an execution throughput that is at least $6\times$ slower than the optimized \texttt{kblas} reference library, highlighting a pervasive LLM limitation: while they can ensure explicit mathematical fidelity in non-native vector spaces, autonomously bridging the gap to micro-architectural optimality remains elusive.

\subsection{Failure Examples}
\begin{figure}[t]
\includegraphics[width=\linewidth]{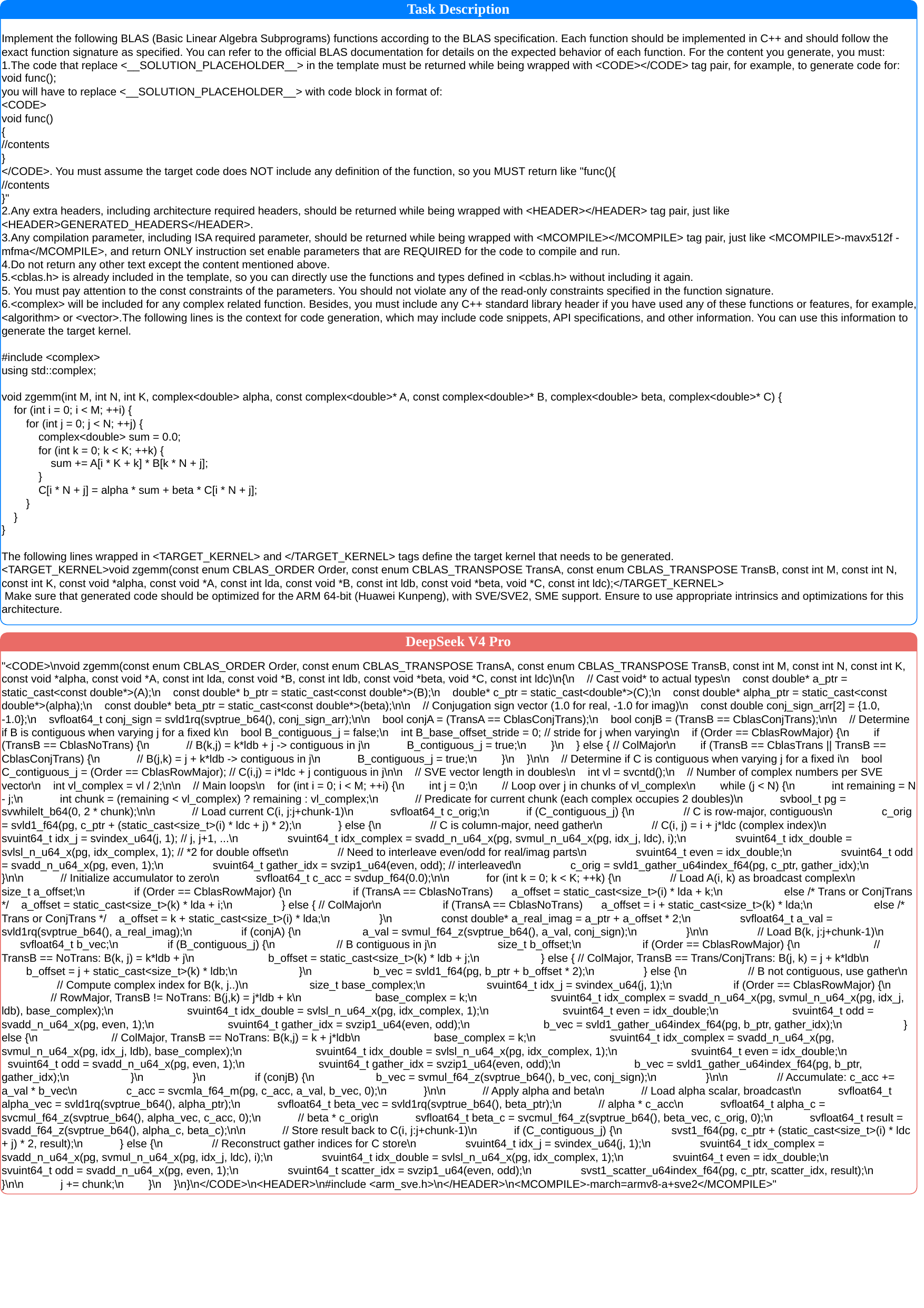}
\caption{Compilation failure in the \texttt{zgemm} case. DeepSeek V4 Pro erroneously invoked \texttt{svcmul\_f64\_z}, an undocumented and non-existent intrinsic, resulting in an immediate compilation abort.}
\label{case_zgemm}
\end{figure}

\paragraph{Failure Example 1:} A fundamentally distinct failure mode, originating from aggressive but misguided hardware targeting, is documented in Fig.~\ref{case_zgemm} for the \texttt{zgemm} operation. In a deliberate attempt to exploit the ARM SVE instructions for peak performance, the DeepSeek V4 Pro model exhibited explicit "hallucination" by invoking a non-existent vector intrinsic function: \texttt{svcmul\_f64\_z}.

Because this fabricated vector instruction is entirely absent from the official ARM C Language Extensions (ACLE), its presence predictably triggered an immediate and fatal compiler abort. This particular failure vector suggests that while the LLM successfully internalizes the localized syntactic conventions of SVE intrinsics such as correctly prefixing with \texttt{sv}, appending the double-precision float type \texttt{\_f64}, and appending a zeroing predication suffix \texttt{\_z}, it critically lacks a rigorously grounded, verifiable mapping of the valid instruction space available for complex operations. This example underscores a profound constraint in current code-generation architectures: the tendency to extrapolate syntactically plausible but semantically non-existent API calls, definitively terminating the compilation pipeline prior to any executable evaluation.

\begin{figure}[t]
\includegraphics[width=\linewidth]{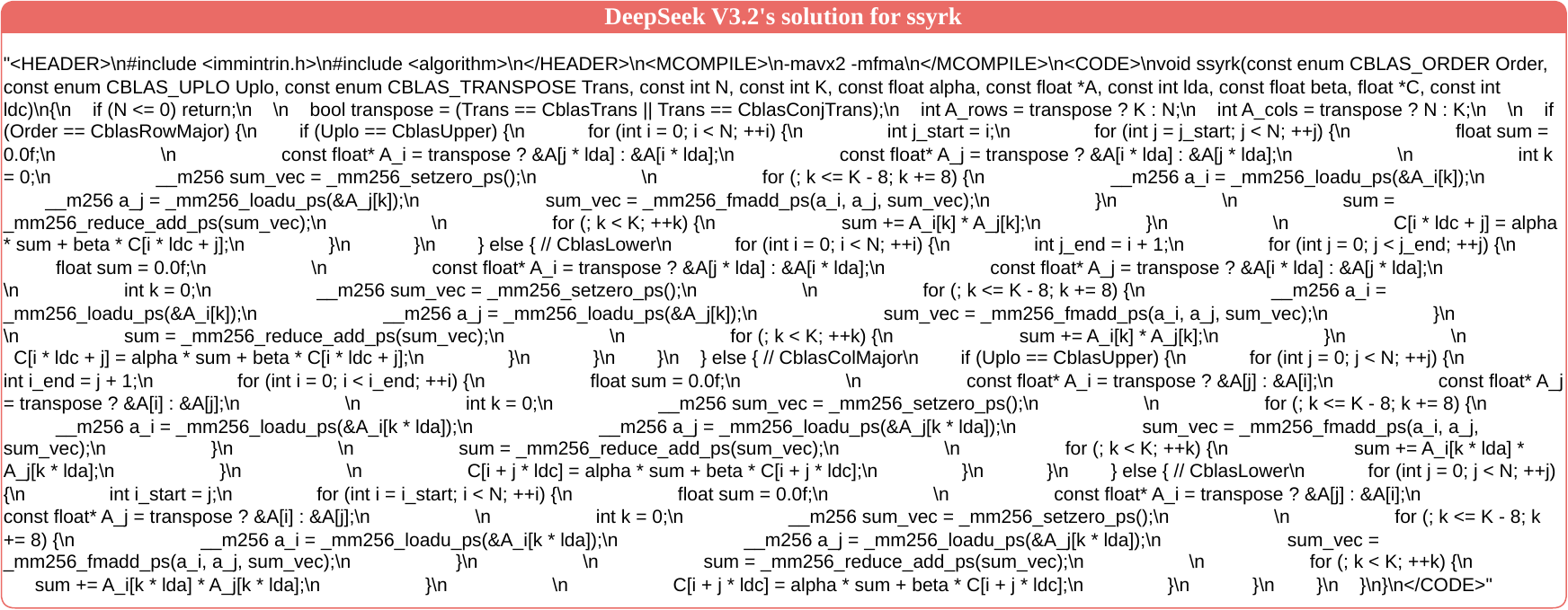}
\caption{Compilation failure in the \texttt{ssyrk} case. DeepSeek V3.2 erroneously invoked \texttt{\_mm256\_reduce\_add\_ps}, an undocumented and non-existent intrinsic, resulting in an immediate compilation abort.}
\label{case_v3.2_ssyrk}
\end{figure}

\begin{figure}[t]
\includegraphics[width=\linewidth]{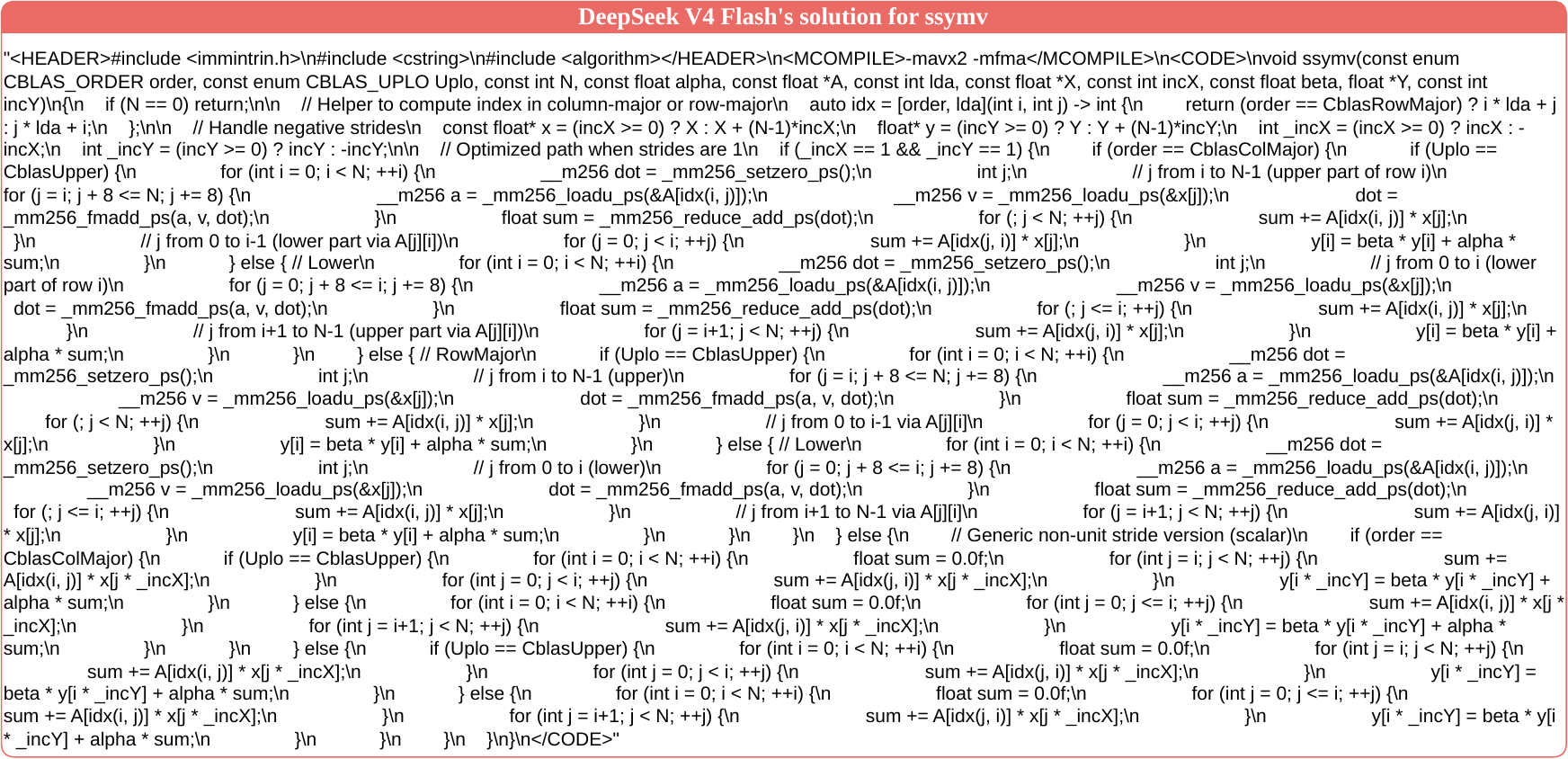}
\caption{Compilation failure in the \texttt{ssymv} case. DeepSeek V4 Flash erroneously invoked \texttt{\_mm256\_reduce\_add\_ps}, an undocumented and non-existent intrinsic, resulting in an immediate compilation abort.}
\label{case_v4flash_ssymv}
\end{figure}

\begin{figure}[t]
\includegraphics[width=\linewidth]{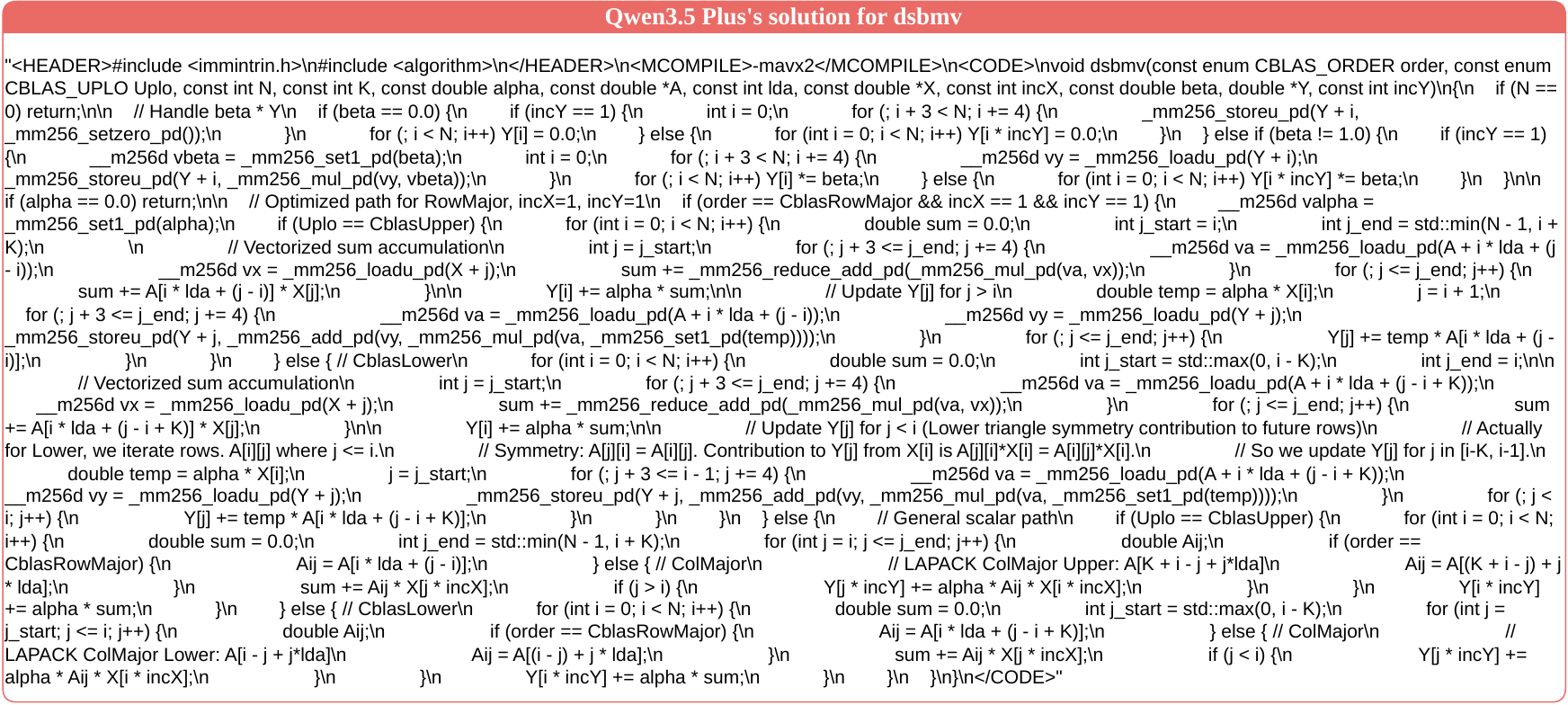}
\caption{Compilation failure in the \texttt{dsbmv} case. Qwen 3.5 Plus erroneously invoked \texttt{\_mm256\_reduce\_add\_pd}, an undocumented and non-existent intrinsic, resulting in an immediate compilation abort.}
\label{case_qwen3.5-plus_dsbmv}
\end{figure}

\begin{figure}[t]
\includegraphics[width=\linewidth]{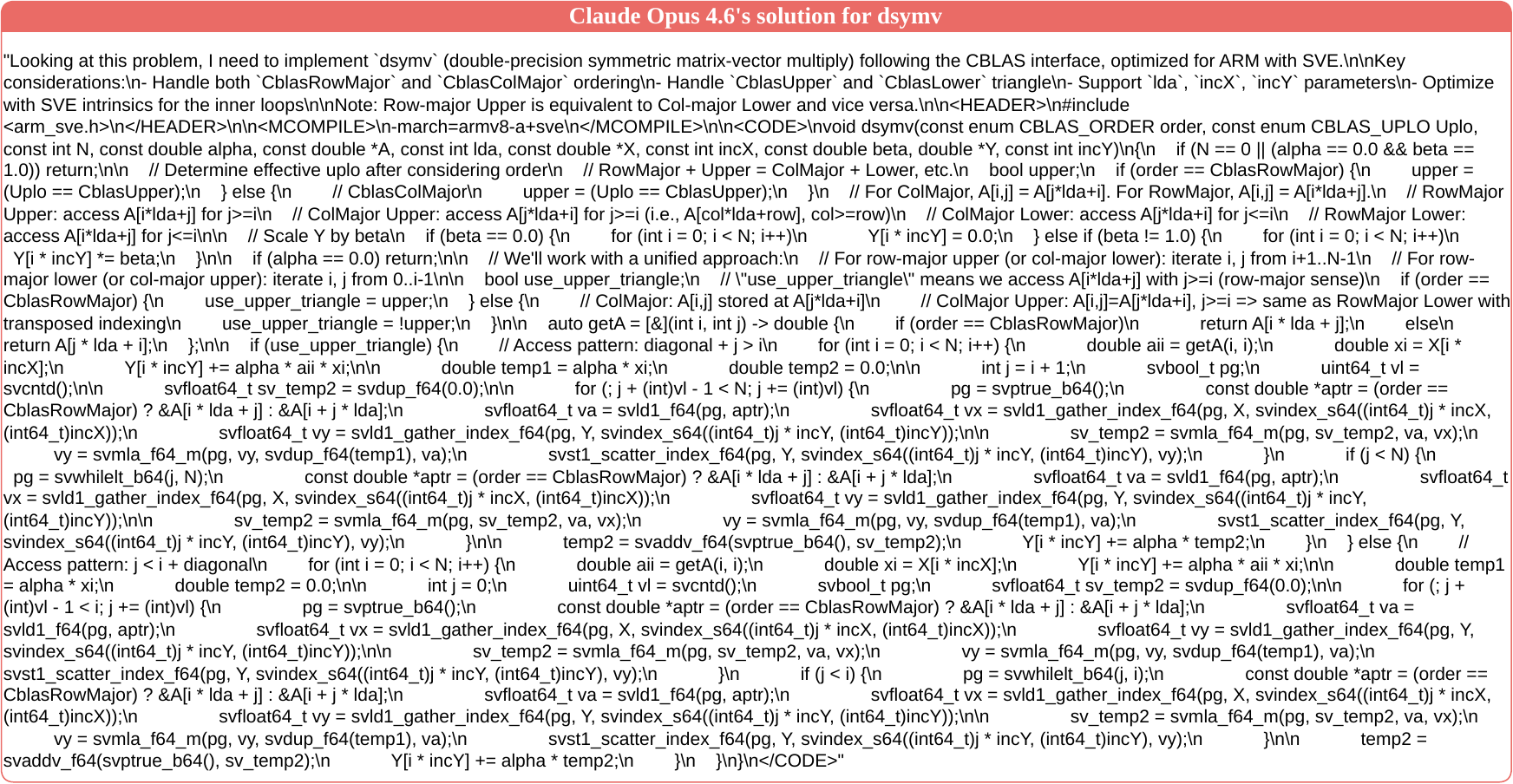}
\caption{Compilation failure in the \texttt{dsymv} case. Claude Opus 4.6 erroneously invoked \texttt{svst1\_scatter\_index\_f64}, an undocumented and non-existent intrinsic, resulting in an immediate compilation abort.}
\label{case_opus4.6_dsymv}
\end{figure}

\paragraph{Failure Example 2:} We discover a common pattern of LLM hallucination in generating kernels. As illustrated in Fig.~\ref{case_v3.2_ssyrk} and Fig.~\ref{case_v4flash_ssymv}, both models provide undocumented intrinsics, trying to finish a multi-step complex operation within one step. And this pattern is not only discovered in DeepSeek series models, as presented in Fig.~\ref{case_qwen3.5-plus_dsbmv}, Qwen 3.5 Plus also tries to use the aligned version of the same non-exist intrinsic. For those considered "premium" models, same behaviour pattern still exists. As shown in Fig.~\ref{case_opus4.6_dsymv}, Claude Opus 4.6 tries to call \texttt{svst1\_scatter\_index\_f64}. However in ACLE spec, scatter intrinsics require precise description in parameters, like \texttt{svst1\_scatter\_s32index\_f32}. Consider that Opus didn't make the same mistakes as the DeepSeeks and Qwens, it can be concluded that in the training process of models like Opus and Sonnet, rich training data on x86\_64 help programs generated by these models to perform correctly on the platforms. Once shifting the platform to those with less training data like Kunpeng, same phenomenon of hallucination appears again.

\clearpage

\end{document}